\documentclass[aps,prl,reprint,superscriptaddress]{revtex4-2} %
\makeatletter

\newif\ifuserevtex
\newif\ifusetwocolfigs

\@ifclassloaded{revtex4-2}{%
  \userevtextrue
  \@ifclasswith{revtex4-2}{reprint}{%
    \usetwocolfigstrue
  }{%
    \usetwocolfigsfalse
  }%
}{%
  \userevtexfalse
  \usetwocolfigsfalse %
}

\makeatother

\newif\ifshowlinenumbers
\usepackage[T1]{fontenc}         %
\usepackage[utf8]{inputenc}      %
\usepackage[english]{babel}      %

\usepackage{amssymb}             %
\usepackage{mathtools}           %
\usepackage{mathrsfs}           %

\usepackage{dcolumn}             %

\usepackage{graphicx}            %
\graphicspath{{./}}

\usepackage{color,xcolor}        %
\usepackage{soul}                %
\usepackage{tcolorbox}           %
\usepackage{cancel}              %
\usepackage{comment}             %
\usepackage{moreverb}            %
\usepackage[normalem]{ulem}      %

\ifuserevtex
    \newcommand{\mybibliostyle}{apsrev4-2}
\else
    \usepackage{amsthm}
    
    \usepackage[a4paper,margin=2.3cm]{geometry} %
    \usepackage{appendix} %
    
    \usepackage[numbers]{natbib} %
    \newcommand{\mybibliostyle}{unsrtnat} %

    \usepackage{authblk} 
    \usepackage{setspace} %
\fi

\usepackage{pifont}

\usepackage{etoolbox}  %
\usepackage{environ}   %
\usepackage{xparse}
\usepackage{scalerel}

\usepackage[
  colorlinks=true,
  linkcolor=blue,  %
  citecolor=blue,  %
  urlcolor=blue    %
]{hyperref}

\usepackage[capitalise]{cleveref}  %

\usepackage{physics}  %
 \showlinenumberstrue  %

\newif\ifshowallemails
\showallemailsfalse  %

\makeatletter

\def\affillist{}
\newcommand{\myaffil}[2]{%
  \expandafter\def\csname affil@#1\endcsname{#2}%
  \ifuserevtex\else
    \listgadd{\affillist}{#1}%
  \fi
}

\newcommand{\buildaffiliations}{%
  \ifuserevtex\else
    \forlistloop{\emitoneaffil}{\affillist}
  \fi
}
\newcommand{\emitoneaffil}[1]{%
  \expandafter\affil\expandafter[#1]{\csname affil@#1\endcsname}%
}

\newcommand{\orcidlink}[1]{%
  \href{https://orcid.org/#1}{\textsuperscript{\textcolor{green!50!black}{\textbullet}}}%
}

\NewDocumentCommand{\myauthor}{O{} O{} O{} m m}{%
  \ifuserevtex
    \author{#4%
      \ifx&#2&\else\ \orcidlink{#2}\fi
    }%
    \ifx&#1&\else
      \ifshowallemails
        \email{#1}%
      \else
        \ifx&#3&\else  %
          \email{#1}%
        \fi
      \fi
    \fi
    \@for\aid:=#5\do{%
      \expandafter\affiliation\expandafter{\csname affil@\aid\endcsname}%
    }%
  \else
    \author[#5]{#4%
      \ifx&#2&\else\ \orcidlink{#2}\fi
      \ifx&#1&\else
        \ifx&#3&%
          \ifshowallemails
            \thanks{\href{mailto:#1}{#1}}%
          \fi
        \else
          \thanks{Corresponding author: \href{mailto:#1}{#1}}%
        \fi
      \fi
    }%
  \fi
}

\newcommand{\storedabstract}{}  %
\newcommand{\storedkeywords}{}  %

\newcommand{\myabstract}[1]{%
  \long\def\storedabstract{#1}%
}

\newcommand{\maketitleandabstract}{%
  \ifuserevtex
    \begin{abstract}
    \storedabstract
    \end{abstract}
    \keywords{\storedkeywords}
    \maketitle
  \else
    \maketitle
    \section*{Abstract}
    \storedabstract
    \ifx\storedkeywords\@empty
    \else
      \vspace{1em}
      \par\noindent \textbf{Keywords:} \storedkeywords
    \fi
  \fi
}

\newcommand{\storedacknowledgements}{}  %

\newcommand{\myacknowledgements}[1]{%
  \gdef\storedacknowledgements{#1}%
}

\newcommand{\insertacknowledgements}{%
  \ifx\storedacknowledgements\@empty
  \else
    \ifuserevtex
      \begin{acknowledgments}
      \storedacknowledgements
      \end{acknowledgments}
    \else
      \section*{Acknowledgements}
      \storedacknowledgements
    \fi
  \fi
}

\makeatother

\ifshowlinenumbers
  \usepackage{lineno}
  \AtBeginDocument{%
    \ifuserevtex
      \ifusetwocolfigs
      \else
        \linenumbers
      \fi
    \else
      \linenumbers
    \fi
  }
\fi

\makeatletter

\def\onecolfig{\@ifnextchar[{\onecolfig@opt}{\onecolfig@opt[]}}
\def\endonecolfig{\end{figure}}
\def\onecolfig@opt[#1]{\begin{figure}[#1]}

\def\twocolfig{\@ifnextchar[{\twocolfig@opt}{\twocolfig@opt[]}}
\def\endtwocolfig{\ifusetwocolfigs \end{figure*} \else \end{figure} \fi}
\def\twocolfig@opt[#1]{%
  \ifusetwocolfigs
    \begin{figure*}[#1]%
  \else
    \begin{figure}[#1]%
  \fi
}

\makeatother

\usepackage{xr}

\makeatletter
\newcommand*{\addFileDependency}[1]{%
  \typeout{(#1)}
  \@addtofilelist{#1}
  \IfFileExists{#1}{}{\typeout{No file #1.}}
}
\makeatother

\newcommand{\papertitle}{Hydrogel mechanics below swelling equilibrium}

\begin{document}

\title{\papertitle}
\myaffil{1}{Institute for Building Materials, ETH Zurich, Switzerland}
\myaffil{2}{Department of Materials, ETH Zurich, Switzerland}

\myauthor[achaocorreas@ethz.ch][0000-0003-0487-7638]{A. Chao Correas}{1}
\myauthor[yanxia.feng@mat.ethz.ch][0009-0004-7866-5747]{Y. Feng}{2}
\myauthor[robert.style@mat.ethz.ch][0000-0001-5305-7658][yes]{R. W. Style}{2}
\myauthor[dkammer@ethz.ch][0000-0003-3782-9368][yes]{D. S. Kammer}{1}

\buildaffiliations \date{\today}

\myabstract{
    Hydrogels are versatile soft materials whose behavior arises from complex polymer-network--fluid interactions. Combining theory and experiments, we show that their mechanics simplify greatly below swelling equilibrium. In this regime, mixing and pore pressures effectively control local hydration, relegating polymer-network elasticity to accommodate the deformed shape. This mechanical hierarchy renders partially-dried hydrogels analogous to incompressible solids with environment-driven target volumes, enabling a staggered mechanical description that accurately captures their state and behavior.
}

\maketitleandabstract

Hydrogels are soft, water-rich polymer networks used in medicine, bioengineering, soft robotics, and advanced manufacturing~\mbox{\cite{Sorrentino2020, Zhou2023, Kim2024, Wang2024, Liu2024, LopezDiaz2024, penn2022optimal}}. In many practical settings, environmental conditions cause their dehydration~\mbox{\cite{Saintyves2023, Yang2024, Liu2025, Webber2025, louf2021under}}. Yet, most characterization and modeling has focused on the near-fully hydrated regime~\mbox{\cite{Durst2011, Sun2012, Agnelli2018, Lei2020, Richbourg2020, Garyfallogiannis2022, Song2025, hu2010using, caccavo2018hydrogels, Mancia2021, Kolvin2018}}, where gels can behave very differently from how they do in drier states~\mbox{\cite{Sekine2009, Xu2023, Xu2024, Sugiyama2024, Li2024, Yerrapragada2024, Zhong2024, Webber2023}}. This raises a broader physical question: if the balance of polymer-network elasticity and polymer--water thermodynamics governs fully swollen gels, are there fundamental changes as they dry below swelling equilibrium?
Here, we show that drying induces a distinct mechanical hierarchy in hydrogels, making them orders of magnitude easier to shear than to undergo drained compression. This occurs because shear is governed by polymer-network elasticity, whereas drained compression requires further water loss and is therefore opposed by the much stronger polymer--water affinity. In fully swollen gels, shape and volumetric deformations are tightly coupled, with local stretching changing water content and vice versa~\cite{li2026following,hu2010using}. Upon dehydration, however, resistance to further water loss strongly stiffens the volumetric response, while polymer-network elasticity often scales only weakly with dehydration~\cite{Flory1943, shao2023independent}. We postulate that this separation of stiffness scales decouples the two deformation modes, simplifying the coupled swelling–-deformation physics into a tiered structure where environmental conditions set shrinkage, which in turn constrains shape deformation. In this sense, drying provides a mechanism through which polymer--water mixing thermodynamics effectively enforces incompressibility on an otherwise compressible material.

\begin{onecolfig}[b]
  \centering
  \vspace{-4.5mm}
  \includegraphics[width=\linewidth]{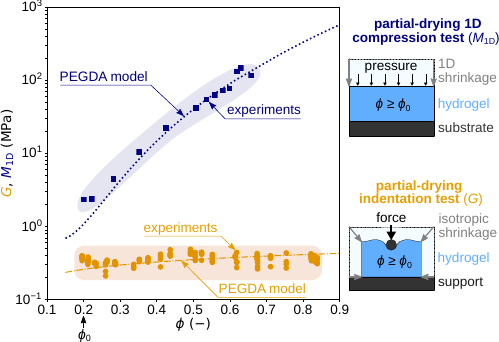}
  \vspace{-5mm}
  \caption{Drying PEGDA hydrogels makes them much easier to undergo shear than drained compression. Blue (orange) points show compression (shear) modulus as a function of polymer content, $\phi$ (reprocessed results from \cite{Feng2025}); schematics illustrate experimental setups. Curves show predictions of the poroelastic model defined in Eqs.~\eqref{eq:1}--\eqref{eq:3} (see SM~\cite{SM}).}
  \label{fig:1}
  \vspace{-2mm}
\end{onecolfig}

We first show how drastically the shear and drained compressive moduli diverge as gels dry out. Using published experimental data for poly(ethylene glycol) diacrylate (PEGDA) hydrogels ($700~\mathrm{g/mol}$) with an as-prepared polymer volume fraction ${\phi_{0}=20\%}$~\cite{Feng2025}, we compare the evolution of the (drained) compressive and shear moduli ($M_{\mathrm{1D}}$ and $G$, respectively) across polymer volume fractions $\phi$ (Fig.~\ref{fig:1}; details in Supplemental Material, SM~\cite{SM}). As PEGDA dries, $M_{\mathrm{1D}}$ increases from~${\sim\!1\,\mathrm{MPa}}$ to~${\sim\!100\,\mathrm{MPa}}$, while $G$ remains relatively constant at~${\sim\!0.1\,\mathrm{MPa}}$. Beyond the specific dependency scalings, the core physical takeaway is clear: dehydration renders hydrogels volumetrically stiff but comparatively compliant against shape changes (${M_{\mathrm{1D}} \gg G}$). Unlike the well-known transient volumetric stiffening due to fluid trapping, this effect is an intrinsic thermodynamic equilibrium property. Analogous results for polyacrylamide and poly(vinyl acetate) gels demonstrate that this behavior is not unique to PEGDA (\cite{Feng2025, Horkay1989}, details SM~\cite{SM}). 

To quantitatively interpret this separation of stiffness scales, we start from the foundational poroelastic description of stresses in hydrogels (\mbox{\cite{Flory1943, Hong2008, Bertrand2016}}; details in SM~\cite{SM}). The corresponding total Cauchy stress tensor, $\underline{\underline{\sigma}}$, represents the bulk stress that is in equilibrium with the external mechanical loading~\cite{Hong2008}:
\begin{equation}
\label{eq:1}
\underline{\underline{\sigma}} = \underline{\underline{\sigma}}_{\mathrm{el}} - \left( \Pi_{\mathrm{mix}} + p \right)\underline{\underline{I}} ~ .
\end{equation}
\noindent Internally, $\underline{\underline{\sigma}}$ is additively decomposed into the elastic stress, $\underline{\underline{\sigma}}_{\mathrm{el}}$, which resists polymer network deformation, the mixing pressure, $\Pi_{\mathrm{mix}}$, reflecting the polymer–water affinity, and the pressure of the interstitial water, $p$. This last is proportional to the chemical potential of the water in the gel (see SM~\cite{SM}) and is generally set by ambient conditions via thermodynamic equilibrium at the outer boundaries (e.g.,~\cite{chen2022nonlinear}). $\underline{\underline{\sigma}}_{\mathrm{el}}$ and $\Pi_{\mathrm{mix}}$ are functions of the displacement gradient, $\nabla\underline{u}$, and the polymer content, ${\phi}$, respectively. However, these two state variables are not independent: assuming intrinsic incompressibility (gel shrinkage occurs only due to water loss), relative local changes in volume are given by ${J = \mathrm{det}\!\left(\underline{\underline{I}}+ \nabla\underline{u}\right)=\phi_{0}/\phi}$~\cite{Bertrand2016}. Thus, we can write Eq.~\eqref{eq:1} as ${\underline{\underline{\sigma}} + p\underline{\underline{I}} = \underline{\underline{\sigma}}_{\mathrm{el}}(\nabla\underline{u}) - \Pi_{\mathrm{mix}}(\nabla\underline{u})\underline{\underline{I}}}$, highlighting that both deformation and dehydration can be driven by changes to either $\underline{\underline{\sigma}}$ or $p$. For example, shrinkage can be achieved by drying a hydrogel (${p<0}$) or through mechanical compression (${\mathrm{tr}(\underline{\underline{\sigma}})<0}$). Indeed, imposing ${p=p_{1}\neq0}$ on a stress-free hydrogel (${\underline{\underline{{\sigma}}} = \underline{\underline{0}}}$) is identical to applying ${\underline{\underline{\sigma}} = p_{1}\underline{\underline{I}}}$ while maintaining ${p=0}$.

We can use Eq.~\eqref{eq:1} to understand the separation of stiffness scales in Fig. \ref{fig:1}. Upon shearing, mechanical response is dictated solely by the polymer-network elasticity, as this sets $G$, which in turn controls the shear response that arises from $\underline{\underline{\sigma}}_{\mathrm{el}}$. The nearly constant $G$ in Fig.~\ref{fig:1} indicates that the network elasticity in PEGDA changes little with dehydration. By contrast, the compressive response involves summed contributions from both network elasticity (via $\underline{\underline{\sigma}}_{\mathrm{el}}$) and polymer-water affinity (via $\Pi_{\mathrm{mix}}$). In particular, since Eq.~\eqref{eq:1} is additive, ${M_{\mathrm{1D}} = M_{\mathrm{1D,\, el}}+M_{\mathrm{1D,\, mix}}\sim G + M_{\mathrm{1D,\,mix}}}$ (${M_{\mathrm{1D,\, el}}\sim G}$ derives from standard polymer network elasticity~\cite{Hong2008}). Given that ${M_{\mathrm{1D}}}$ increases by orders of magnitude with drying while $G$ remains flat, the compressive stiffening is dominated by the mixing contribution $M_{\mathrm{1D,\, mix}}$; specifically, by the increasing difficulty of stripping water molecules from a drying, hydrophilic hydrogel. This scale separation reflects the dominance of polymer–water affinity over polymer-network elasticity in the volumetric mechanics of drying hydrogels (see verification in SM~\cite{SM}).

The emerging mechanical hierarchy in drying hydrogels suggests a profound physical simplification: mixing pressure balances pore pressure almost entirely on its own (${\Pi_{\mathrm{mix}} \approx -p}$). Given the negligible influence of polymer-network elasticity on the volumetric response of dehydrated hydrogels, $\vert\mathrm{tr}(\underline{\underline{\sigma}}_{\mathrm{el}})\vert \ll \vert\Pi_{\mathrm{mix}}\vert$ holds regardless of mechanical loading and deformation (except under uncommon, large triaxial stresses). This simplifies the volumetric stress balance in Eq.~\eqref{eq:1} into ${\Pi_{\mathrm{mix}} \approx -p}$. Crucially, such reduction is not a constitutive artifact; rather, it is grounded in the well-established additive stress decomposition and the observed stiffness scale separation ($M_{\mathrm{1D}} \gg G$). Therefore, the resulting mechanical hierarchy in drying hydrogels is constitutive-agnostic, dictating that ambient-imposed pore pressure sets the polymer volume fraction ${\phi \approx \Pi_{\mathrm{mix}}^{-1}(-p)}$, while the elastic contribution to shrinkage becomes subdominant.

\begin{onecolfig}[b]
  \vspace{-4.5mm}
  \centering
  \includegraphics[width=\linewidth]{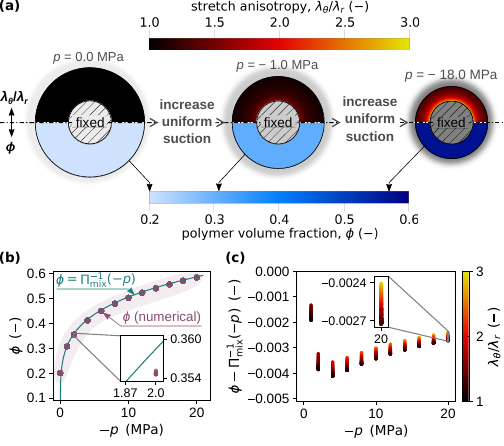}
  \vspace{-7mm}
  \caption{Shrinkage in hydrogels below swelling equilibrium is set by the balance of mixing and pore pressures, ${\Pi_{\mathrm{mix}} = -p}$, regardless of geometric constraints. Simulations of an annular hydrogel fixed at its inner boundary and subjected to increasingly negative pore pressures~$p$ show: (a)~contours of stretch anisotropy~$\lambda_{\theta}/\lambda_{r}$ and polymer volume fraction~$\phi$; (b)~local $\phi$ versus the equilibrium solution~${\phi = \Pi_{\mathrm{mix}}^{-1}(-p)}$; and (c)~local error ${\phi - \Pi_{\mathrm{mix}}^{-1}(-p)}$, colored by~${\lambda_{\theta}/\lambda_{r}}$.}
  \vspace{-3mm}
  \label{fig:2}
\end{onecolfig}

To quantitatively assess the validity of this leading-order balance, we adopt conventional constitutive laws as a reference model for hydrogel poroelasticity. Because the underlying mechanical hierarchy is expected to be generic to drying hydrogels, other suitable constitutive choices should yield the same qualitative behavior; the specific formulation adopted here is used only to enable explicit quantitative analysis. To this end, we complete Eq.~\eqref{eq:1} by specifying the elastic stress as~\cite{Hong2008, Garyfallogiannis2022}:
\begin{equation}
\label{eq:2}
\underline{\underline{\sigma}}_{\mathrm{el}}\!\left(\nabla \underline{u}\right) = 
\frac{G_{0}}{J} \left[ \underline{\underline{B}} - \underline{\underline{I}} \right] + \sigma_{0} \underline{\underline{I}}~,
\end{equation}
\noindent where $G_{0}$ is the shear modulus in the undeformed configuration, $\underline{\underline{B}}$ is the left Cauchy--Green tensor associated with $\nabla\underline{u}$, and ${\sigma_{0} \underline{\underline{I}}}$ ensures a stress-free undeformed configuration, i.e., ${\sigma_{0} = \Pi_{\mathrm{mix}}\!\left(\phi_{0}\right)}$. For the mixing pressure, we adopt a des Cloizeaux-type power law in $\phi$~\cite{Feng2025}:
\begin{equation}
\label{eq:3}
\Pi_{\mathrm{mix}}\!\left(\phi\right) = 
\frac{A k_{\mathrm{B}} T}{v_{w}}\phi^{n}~,
\end{equation}
\noindent where $A$ is a dimensionless constant that governs the intensity of polymer--water affinity, $k_{\mathrm{B}}$ is the Boltzmann constant, $T$ is the absolute temperature, $v_{w}$ is the molecular volume of water, and $n$ relates to the fractal architecture of the porous polymer network~\cite{Feng2025}. Because the elastic and mixing stress contributions enter the same equation, this poroelastic formulation couples volume change and shape deformations; hence, we refer to this model as the strongly coupled formulation. Using ${G_{0} = 0.26~\mathrm{MPa}}$, ${A = 1.75}$, and ${n = 4.67}$ (calibration details in SM~\cite{SM}), this constitutive model describes the results in Fig.~\ref{fig:1}. Henceforth, we adopt these values and assume plane-strain conditions.

Using the strongly coupled formulation, we confirm that drying hydrogels naturally obey ${\Pi_{\mathrm{mix}} \approx -p}$, even under complex anisotropic deformation. Specifically, we consider an annular hydrogel fixed at its inner boundary and subjected to a homogeneous negative pore pressure, $p$ (Fig.~\ref{fig:2}; details in SM~\cite{SM}). As suction intensifies, the hydrogel dehydrates and shrinks, but the inner constraint prevents isotropic contraction and forces inhomogeneous shape distortion. We quantify deviations from isotropic shrinkage via the stretch anisotropy: the ratio of hoop to radial stretch ${\lambda_{\theta}/\lambda_{r}}$. Even as the hydrogel shape distorts considerably (locally reaching ${\lambda_{\theta}/\lambda_{r} \approx 3}$), ${\phi\! =\!\Pi^{-1}_{\mathrm{mix}}\!\left(- p\right)}$ remains remarkably accurate. Notably, it holds even at mild levels of dehydration (Fig.~\ref{fig:2}b), with minimal sensitivity to stretch anisotropy (Fig.~\ref{fig:2}c). This demonstrates that shape distortions and polymer-network elasticity play negligible roles in determining dehydration-driven shrinkage. Instead, network elasticity is relegated to the subordinate role of determining the shape compatible with mechanical boundary conditions. 

The robustness of ${\phi\! =\!\Pi^{-1}_{\mathrm{mix}}\!\left(- p\right)}$ suggests that drying hydrogels effectively behave as incompressible solids during mechanical deformation, yet with a target volume governed by the environment. Crucially, this is fundamentally different from their well-known transient incompressibility~\cite{Bertrand2016}, where rapid volume changes are delayed by trapped interstitial water; here, the effective  incompressibility is an intrinsic equilibrium feature. This warrants a staggered mechanical description: instead of solving for volume change and shape deformation simultaneously, we first use the ambient-imposed pore pressure to determine the local polymer volume fraction, and then we solve only for the shape compatible with the resulting volume constraint. In this weakly coupled approximation (denoted by hats), the polymer volume fraction is:
\begin{equation}
\label{eq:4}
\hat{\phi} \left(p\right) = \mathrm{max} \!\left[\Pi_{\mathrm{mix}}^{-1}(-p),\, \phi_{0}\right]~,
\end{equation}
\noindent where the $\mathrm{max}$ operator enforces the hydrated reference state at ${p=0}$. The polymer volume fraction $\hat{\phi}(p)$ sets the local relative volume through ${\hat{J} = \phi_{0}/\hat{\phi}(p)}$. The remaining mechanical problem reduces to an elastic deformation problem subject to a volumetric constraint. Taking the elastic stress from Eq.~\eqref{eq:2}, this is defined as:
\begin{equation}
\label{eq:5}
\left\{
\begin{array}{ll}
\underline{\underline{\hat{\sigma}}}\!\left(\nabla\underline{\hat{u}},\, \hat{\pi}\right) = 
\displaystyle \frac{G_{0}}{\hat{J}} \left[ \underline{\underline{\hat{B}}} - \underline{\underline{I}} \right] + \hat{\pi} \underline{\underline{I}}  \\
\displaystyle \mathrm{subject~to~}\hat{J} = \phi_{0}/\hat{\phi} \left(p\right) ~~~~~~~~ ,
\end{array}
\right.
\end{equation}
\noindent which mathematically expresses the analogy to incompressible solids (see e.g.~\cite{Holzapfel2000}). The Lagrange multiplier ${\hat{\pi}\underline{\underline{I}}}$ is not a prescribed pressure but rather a reactive field that forces the elastic deformation to comply with the local volume constraint. As shown in SM~\cite{SM}, ${\vert \hat{\pi}/p\vert}$ serves as a self-contained accuracy metric, where ${\vert \hat{\pi}/p\vert \ll 1}$ indicates near-exact agreement with the strongly coupled formulation. Note that we now define ${\sigma_{0} = 0}$ so that elastic stresses vanish at the reference state (${p=0}$ and ${{\nabla\underline{u}} = \underline{\underline{0}}}$). The strongly coupled mechanics in Eqs.~\mbox{(\ref{eq:1}--\ref{eq:3})} decouples into a sequential hierarchy: Eq.~\eqref{eq:4} sets the ambient-induced shrinkage, and Eq.~\eqref{eq:5} drives the mechanically compatible shape. Far from a heuristic simplification, this formulation mathematically formalizes an already-excellent physical approximation grounded in the observed mechanical hierarchy.

\begin{onecolfig}[b]
  \vspace{-5mm}
  \centering
  \includegraphics[width=\linewidth]{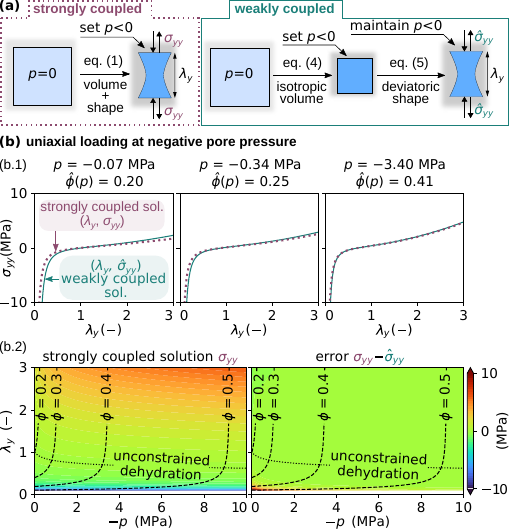}
  \vspace{-6mm}
  \caption{Drying hydrogels deform analogously to incompressible solids with an environment-set target volume. We compare stress predictions from strongly coupled (standard) poroelasticity, ${\sigma_{yy}}$, and weakly coupled environment-driven, pseudo-incompressible elasticity, $\hat{\sigma}_{yy}$, under negative pore pressure ${p \le 0}$ and uniaxial stretch ${\lambda_{y}\in\left[0.2,\, 3\right]}$. (a) Formulations and loading conditions. (b.1) Stress-stretch curves for varying ${p}$. (b.2)~Contours of $\sigma_{yy}$ and the error ${\sigma_{yy} - \hat{\sigma}_{yy}}$ across~$\lambda_{y}$ and~$p$; dashed lines denote iso-hydration curves.}
  \vspace{-2mm}
  \label{fig:3}
\end{onecolfig}

The weakly coupled formulation accurately describes the mechanical state and behavior of hydrogels under combined dehydration and mechanical loading. To show this, we consider a representative element that is simultaneously subjected to uniaxial stretch, $\lambda_{y}$, and negative pore pressure, $p$ (Fig.~\ref{fig:3}a, details in SM~\cite{SM}). The resulting stress--stretch curves show excellent agreement across varying degrees of dehydration (Fig.~\ref{fig:3}b.1). Noticeable deviations are confined to cases of severe uniaxial compression ($\lambda_{y} \rightarrow 0$) combined with weak pore pressure. However, a systematic error analysis confirms that this specific regime is rather narrow (Fig.~\ref{fig:3}b.2). Away from these specific conditions, the stress discrepancy ${\sigma_{yy} - \hat{\sigma}_{yy}}$ is negligible; remarkably, this is true even when $\vert\sigma_{yy}\vert \sim \vert p\vert$. These results broadly validate the weakly coupled formulation for drying hydrogels, even under intense mechanical loading. Beyond its accuracy, this formulation offers a clear mathematical advantage: while the weakly coupled formulation has an analytical solution, the strongly coupled formulation requires an iterative root finder (see SM~\cite{SM}).

\begin{onecolfig}[b]
  \vspace{-5mm}
  \centering
  \includegraphics[width=\linewidth]{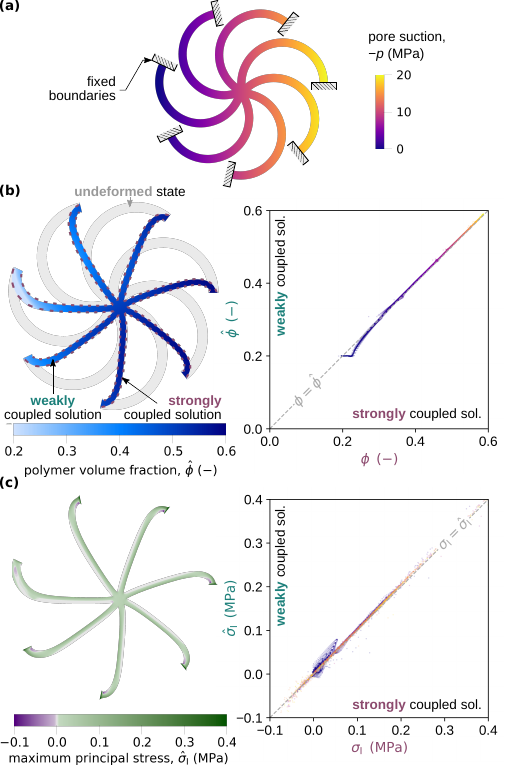}
  \caption{Inhomogeneous hydrogel drying is captured by weakly coupled pseudo-incompressible mechanics, even under large deformations and complex geometry. We consider a chiral spiral with fixed tips under linear pore suction, ${p \propto -X}$. (a)~Problem schematic. (b) Left: strongly coupled poroelastic deformation (dashed outline) over the weakly coupled solution (contours of polymer volume fraction, ${\hat{\phi}}$); right: pointwise comparison of $\phi$ and $\hat{\phi}$, colored by $p$. (c) Corresponding comparison of the maximum principal stress, ${\sigma_{\mathrm{I}}}$ and ${\hat{\sigma}_{\mathrm{I}}}$.}
  \vspace{-2mm}
  \label{fig:4}
\end{onecolfig}

Moving beyond canonical benchmarks, we validate the weakly coupled formulation on hydrogels with complex shapes undergoing extreme drying-induced deformation. We numerically simulate a chiral seven-arm spiral with fixed tips subjected to nonhomogeneous pore suction, $p$ (Fig.~\ref{fig:4}a; details in SM~\cite{SM}). 
This benchmark is particularly stringent: the chirality triggers a constrained unwinding motion during drying, causing large deformations and rotations. Despite these geometric nonlinearities, both formulations predict nearly identical deformed outlines. Quantitative node-wise comparisons of the polymer volume fraction (Fig.~\ref{fig:4}b) and maximum principal stress (Fig.~\ref{fig:4}c) confirm their agreement. Deviations are minor and confined to regions with weak pore suction ($p\!\rightarrow\!0$), corresponding to nearly fully hydrated states ($\{\phi,\, \hat{\phi}(p)\}\!\rightarrow\!\phi_{0}$). 

In conclusion, our results reveal a mechanical hierarchy in drying hydrogels: it is much easier to change their shape than their volume by further expelling water. This contrast stems from a sharp separation of stiffness scales upon drying, wherein the strong polymer--water mixing pressure yields high volumetric stiffness, while polymer-network elasticity governs the relatively weaker resistance to shape deformation. Consequently, local hydration depends almost exclusively on the chemical potential of the interstitial water, virtually independent of the applied mechanical loads. Drying thus renders hydrogels analogous to incompressible solids, but with an environment-driven target volume—showing how mixing thermodynamics translates into effective mechanical constraints in gels. This physical insight allows us to recast hydrogel mechanics below swelling equilibrium into a weakly coupled continuum description that accurately captures their overall response and internal mechanical state.

We expect this mechanical hierarchy to apply broadly to drying hydrogels, rather than being specific to the constitutive laws considered here. Because of their inherent affinity for water, the volumetric stiffness of hydrogels should always increase by orders of magnitude upon drying, transitioning from soft, swollen gels to dry polymers \cite{bhattacharyya2020hydrogel,ito2025osmotic}. By contrast, polymer-network physics predicts only weak scaling of the shear modulus with polymer content (${G\propto\phi^{1/3}}$) \cite{shao2023independent}, implying that shear stiffness is generally much smaller than volumetric stiffness at low water contents. Exceptions may arise in gels with strongly nonlinear elastic networks (e.g., many biological tissues \cite{yan2025water,xu2023synthetic,wahlsten2023multiscale}), or those undergoing glass transition \cite{delavoipiere2016poroelastic}, crystallization \cite{lin2019anti}, or stimuli-responsive changes \cite{neumann2023stimuli} during dehydration. Future work should test the universality of this hierarchy across hydrogel chemistries under different environmental and mechanical conditions.

Our physical insights extend beyond theory and have several functional implications for hydrogel research. They provide a simplified viewpoint for understanding and modeling stresses and deformations during drying or swelling, with applications for predicting cracking stresses in drying gels and epithelial tissue \cite{german2012heterogeneous}, swelling instabilities \cite{trujillo2008creasing,Bertrand2016}, and volume changes in hydrogel sorbents \cite{kabiri2011superabsorbent}; designing hydrogel actuators with pre-programmed shape changes \cite{ionov2014hydrogel,moser2022hydroelastomers,shen2026programmable}, and understanding load bearing in hydrated tissues such as cartilage \cite{cederlund2022walking, spencer2014aqueous}. Practically, they suggest new sensing applications: because gel volume only responds to variations in water chemical potential but is nearly insensitive to mechanical loading, volume changes can provide load-invariant measurements of humidity, temperature, and pore pressure \cite{zhou2024review}, even in mechanically active or biological environments. 

\myacknowledgements{
\textit{Acknowledgments ---} ACC and DSK acknowledge the funding received from the ETH Zurich Postdoctoral Fellowship programme through the \mbox{CryoCracks} research project. RWS and YF acknowledge support from the Swiss National Science Foundation (200021-212066).
A. Chao Correas thanks $\mathrm{J.\,\, A. \,\, Chao \,\, Huetos^{\dagger}}$ for his lifelong support and guidance.
}
\insertacknowledgements

\textit{Code \& data availability ---} 
The code used for the numerical simulations is available on \href{https://gitlab.ethz.ch/smec/papers-supp-info/2026/hydrogel-mechanics-below-swelling-equilibrium}{ETH GitLab} and the generated data has been deposited in the ETH Research Collection.

\textit{Note added ---} The Authors declare the use of AI-powered tools to improve readability.

\bibliographystyle{\mybibliostyle}
%
%
%
%
%
%
%
%
%
%
 %

%
\clearpage
\appendix
\section{\textsc{Supplemental Material}}
\section{A. Experimental characterization of the separation of stiffness scales in gels below swelling equilibrium}

In this section, we demonstrate with published experimental data that the separation of stiffness scales is a generic characteristic of polymeric gels, rather than a feature unique to specific chemistries. We divide this empirical analysis into two blocks. First, we analyze hydrogel systems with water as a solvent—specifically, poly(ethylene glycol) diacrylate~(PEGDA) and polyacrylamide~(PAM)—drawing on measurements originally published in Ref.~\cite{Feng2025}. Second, we complement these observations with independent results on poly(vinyl acetate) organogels with different organic solvents, originally published in Ref.~\cite{Horkay1989}.

The specific hydrogels examined in the first part (Sects.~A.1 and A.2) are PEGDA (molar mass $700~ \mathrm{g/mol}$ and initial polymer volume fraction $\phi_{0} = 20~\%$) and PAM ($\phi_{0} = 9.57~\%$). For the second block (Sects.~A.3 and A.4), we examine poly(vinyl acetate) networks imbued with toluene and acetone (PVAc-t and PVAc-a, respectively). For both organogel chemistries, we restrict our analysis to the published results for $9\%$ initial polymer weight fraction; specifically, we consider polymer-to-crosslinker ratios of ${100:1}$, ${200:1}$, and ${400:1}$, as only these chemistries are available for both solvents. The additional gel formulations reported in both cited references yield analogous results. In what follows, we briefly outline the experimental techniques and postprocessing steps applied to these literature datasets.

\subsection{A.1. One-dimensional compressive modulus $M_{\mathrm{1D}}$ under partial dehydration $\phi \ge \phi_{0}$ via gel-freezing osmometry}

We characterize the volumetric stiffness as a function of the polymer volume fraction $\phi$ from gel freezing osmometry tests~\cite{Feng2025}. In this technique, a hydrogel sample is brought into contact with ice. As the system cools below the melting temperature of water $T_{\mathrm{m}}$, ice draws the interstitial water out of the hydrogel via cryosuction. At the \mbox{ice--hydrogel} boundary, the pore pressure $p$ is prescribed by the Clausius--Clapeyron relation:
\begin{equation}
\label{eq:sm.a1.1}
p\left(T\right)= - \rho_{\mathrm{w}}L_{\mathrm{w}}\frac{T_{\mathrm{m}}-T}{T_{\mathrm{m}}} \, ,
\end{equation}
\noindent where $\rho_{\mathrm{w}}$ and $L_{\mathrm{w}}$ represent the density and latent heat of fusion of water, respectively, and $T$ is the local absolute temperature. Given sufficient time and a homogeneous ice temperature, this boundary pressure equilibrates throughout the bulk. Per standard hydrogel poroelasticity (see Eq.~\eqref{eq:1}), this state is mathematically equivalent to an isotropic compression test driven by the homogeneous hydrostatic pressure $p(T)$. 

The referenced experiments use flat hydrogel films with an initial thickness of ${h_{0} \approx 150~\mu\mathrm{m}}$ and much larger in-plane dimensions. Because the samples are adhered to a glass substrate on one face and are in contact with ice on the other, their deformation is effectively one dimensional, confining shrinkage to occur along the thickness direction. Let $\lambda = h/h_{0}$ be the one-dimensional stretch, where $h$ is the current thickness. The volume change is then ${J = \lambda}$, and the polymer volume fraction is ${\phi = \phi_{0}/\lambda}$. The resulting one-dimensional compressive modulus $M_{\mathrm{1D}}$, which acts as an experimental proxy for the volumetric stiffness, is defined as: 
\begin{equation}
\label{eq:sm.a1.2}
M_{\mathrm{1D}}= \lambda\frac{\mathrm{d}p}{\mathrm{d}\lambda} \, .
\end{equation}

By setting a target temperature and measuring thickness at equilibrium, these experiments yielded discrete pairs of temperature $T$ and thickness $h$. We convert these pairs into the corresponding values of $p$, $\lambda$, and $\phi$. To evaluate the derivative in Eq.~\eqref{eq:sm.a1.2} from these discrete data, we apply finite differences using the \texttt{numpy.gradient} function. The resulting ${M_{\mathrm{1D}}(\phi)}$ estimates for PEGDA and PAM are plotted in Figs.~\ref{fig:1} and \ref{fig:sm.a.1}, respectively.
\begin{onecolfig}[t]
  \centering
  \includegraphics[width=\linewidth]{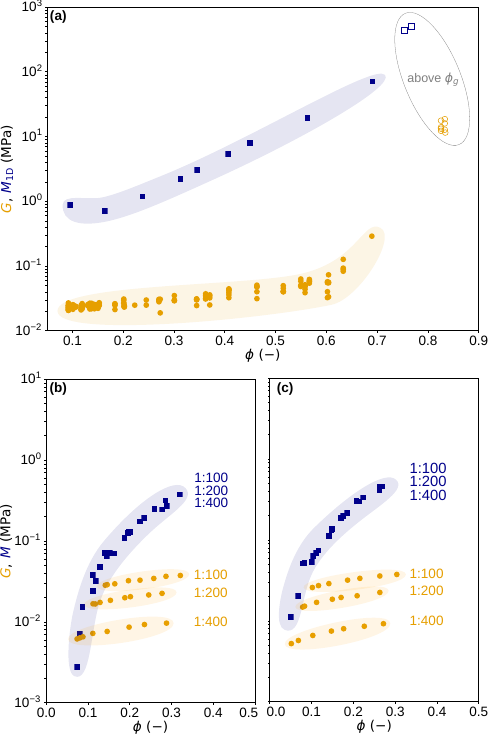}
  \caption{The separation of stiffness scales is also observed in PAM hydrogels and PVAc organogels below swelling equilibrium. Experimental results at various degrees of drying/deswelling for: (a)~ one-dimensional compressive modulus $M_{\mathrm{1D}}$ and apparent shear modulus $G$ of PAM~\cite{Feng2025} (PAM undergoes a glass transition for ${\phi \ge \phi_{g} \approx 0.73}$); (b,~c)~bulk modulus $M$ and shear modulus $G$ of PVAc-t and PVAc-a organogels~\cite{Horkay1989}, respectively. Crosslinker-to-polymer molar ratios are indicated on the right of each curve.}
  \label{fig:sm.a.1}
\end{onecolfig}
\subsection{A.2. Apparent shear modulus $G$ under partial dehydration $\phi \ge \phi_{0}$ via microindentation}

We rely on the experimental characterization of the shear modulus $G$ as a function of $\phi$ reported in Ref.~\cite{Feng2025}. For clarity, we briefly summarize the experimental protocol and data processing. First, cylindrical hydrogel samples of radius ${\approx 30~\mathrm{mm}}$ and height ${\approx 3~\mathrm{mm}}$ are allowed to dehydrate unconstrained, shrinking volumetrically such that ${J = \phi_{0}/\phi}$. Then, they are subjected to indentation tests with a hemispherical indenter of radius ${R_{\mathrm{i}} = 1.58~\mathrm{mm}}$. Because the indentation duration was short enough to preclude significant water migration, samples remained locally incompressible, i.e. ${J = \phi_{0}/\phi}$ throughout indentation. Adopting the corresponding Hertzian contact model for a semi-infinite elastic domain, the apparent shear modulus $G$ is: 
\begin{equation}
\label{eq:sm.a2.1}
G = \frac{3}{16} \frac{F_{\delta}}{\sqrt{R_{\mathrm{i}} \, \delta^{3}}}  \, ,
\end{equation}
\noindent where $F_{\delta}$ is the force required to reach an indentation depth $\delta$. For each dehydration state, ${G}$ is obtained from Eq.~\eqref{eq:sm.a2.1} using the force at the largest experimental indentation (${\delta \sim 0.1~\mathrm{mm}}$). For highly dehydrated samples whose reduced thickness does not warrant the semi-infinite domain assumption, finite size corrections were used (see details in the Supplemental Materials of Ref.~\cite{Feng2025}). The resulting ${G(\phi)}$ values for PEGDA and PAM are plotted in Figs.~\ref{fig:1} and \ref{fig:sm.a.1}, respectively. Notably, the data for PAM reveal a sharp increase in $G$ at high polymer volume fractions. As reported in Ref.~\cite{Feng2025}, this constitutive nonlinearity corresponds to a glass transition occurring at ${\phi \ge \phi_{g} \approx 0.73}$. 
\subsection{A.3. Bulk modulus $M$ under partial deswelling $\phi \ge \phi_{0}$ via the osmotic stress technique}

We characterize the volumetric stiffness of PVAc organogels as a function of $\phi$ using published osmotic deswelling data~\cite{Horkay1989}. This technique involves enclosing a gel sample in a semipermeable membrane and immersing it in a bath containing a solution of linear, uncrosslinked PVAc in the matching solvent (toluene or acetone). The osmotic suction exerted by the solution draws solvent out of the gel across the membrane. In our notation, this osmotic suction corresponds to a negative pressure ${p \le 0}$, the magnitude of which is determined by the concentration of PVAc monomers in the external solution ${\phi_{\mathrm{sol}}}$. The functional relationship ${p(\phi_{\mathrm{sol}})}$ for the considered polymer-solvent combinations is known from independent measurements~\cite{Vink1974}.

The cited experiments were carried out on cylindrical PVAc samples with equal diameter and height (${\approx 1\,\mathrm{cm}}$). By varying ${\phi_{\mathrm{sol}}}$, the samples deswelled until thermodynamic equilibrium was attained. They were then weighed to measure solvent loss and determine the resulting polymer volume fraction. Based on the reported values of ${\phi}$ as a function of $p$, we determine the bulk modulus $M$ as:
\begin{equation}
\label{eq:sm.a3.1}
M = -\phi \frac{\mathrm{d} p}{\mathrm{d}\phi},
\end{equation}
where the derivative is evaluated using the \texttt{numpy.gradient} function, analogous to Sect.~A.1. The resulting ${M(\phi)}$ estimates for PVAc-t and PVAc-a are plotted in Fig.~\ref{fig:sm.a.1}~(b,~c), respectively. The lack of dependence of $M$ on the crosslinker content in the deswollen state (${\phi \ge \phi_{0}}$) demonstrates that it is dominated by the mixing contribution (${M \approx M_{\mathrm{mix}}}$), as already postulated in the main text. This is because crosslinker content affects only the formation of the polymer network, not the interactions between the polymer and the solvent.
\subsection{A.4. Shear modulus $G$ under partial deswelling $\phi \ge \phi_{0}$ via uniaxial compression}

We rely on the experimental characterization of the shear modulus $G$ as a function of $\phi$ for PVAc organogels reported in Ref.~\cite{Horkay1989}. As in Sect.~A.2, we briefly summarize their experimental protocol and data processing for clarity. Values for the shear modulus at different degrees of deswelling were determined by subjecting the cylindrical samples (${\phi \ge \phi_{0}}$) to unidirectional compression. The loading timescale (${\mathcal{O}(10^{2}\ \mathrm{s})}$) was much shorter than the characteristic deswelling time (${\mathcal{O}(10^{5}\ \mathrm{s})}$), forcing the sample to deform isochorically. Over the range of axial compression investigated (${\lambda\in\left[0.7,\, 1\right]}$, with ${\lambda = h/h_{0}}$ as the axial stretch), the force–stretch response naturally followed neo-Hookean elasticity. Hence, the shear modulus $G$ is given by:
\begin{equation}
\label{eq:sm.a4.1}
G =\frac{F_{\lambda}}{{A_{\mathrm{nom}}}} \left(\frac{\lambda^{2}}{\lambda^{3} -1}\right)
\end{equation}
where ${F_{\lambda}}$ is the compressive force and ${A_{\mathrm{nom}}}$ is the resisting area of the deswollen sample prior to mechanical loading. The absence of barreling and volume changes was experimentally verified. The resulting ${G(\phi)}$ estimates for PVAc-t and PVAc-a are plotted in Fig.~\ref{fig:sm.a.1}~(b,~c), respectively.

\section{B. Strongly coupled poroelastic formulation for hydrogel mechanics}

We employ a Lagrangian continuum framework to describe standard poroelasticity in hydrogel mechanics. This formulation is strongly coupled in a constitutive sense, meaning the volume and shape responses are intrinsically interdependent. While the Lagrangian description is mathematically advantageous because it naturally tracks the affine deformation of the polymer network, we analyze our results in the main text using Eulerian quantities, as they represent the actual physical state of the hydrogel. Furthermore, we restrict our primary scope to the stationary limit of poroelasticity, focusing on the mechanical steady-state, where pore pressure is assumed to be time invariant and known. Below, we provide the corresponding kinematics, constitutive equations, governing principles, material characterization, and analytical particularizations.%

\subsection{B.1. Lagrangian kinematics and mass conservation}

Let ${\Omega_{0} \subseteq \mathbb{R}^{N}}$ be an $N$-dimensional continuum domain representing the reference (undeformed) configuration of the hydrogel. We identify the material points of the polymeric network using the position vector field ${\underline{X}\in\Omega_{0}}$. Under an affine deformation defined by the network displacement field ${\underline{u}: \Omega_{0} \rightarrow \mathbb{R}^{N}}$, the domain maps to its (deformed) current configuration ${\Omega \subseteq \mathbb{R}^{N}}$ via the kinematic relation ${\underline{x}\!\left(\underline{X}\right) = \underline{X} + \underline{u}\!\left(\underline{X}\right)}$. This yields the deformation gradient tensor $\underline{\underline{F}}$:
\begin{equation}
\label{eq:sm.b1.1}
\underline{\underline{F}} =  
\frac{\partial \underline{x}}{\partial \underline{X}} = 
\underline{\underline{I}} + \nabla\underline{u}\, ,
\end{equation}
\noindent where $\nabla\underline{u}$ is the gradient of $\underline{u}$ with respect to the reference (material) coordinates $\underline{X}$. The local volumetric change following deformation is given by the determinant of $\underline{\underline{F}}$, namely $J = \mathrm{det}\!\left(\underline{\underline{F}}\right)$. Based on ${\underline{\underline{F}}}$, we define the right Cauchy-Green deformation tensor $\underline{\underline{C}}$:
\begin{equation}
\label{eq:sm.b1.2}
\underline{\underline{C}} = \underline{\underline{F}}^{\mathrm{T}} \underline{\underline{F}}\, ,
\end{equation}
\noindent whose eigenvalues are the squares of the principal stretches, i.e., ${\lambda_{i}^{2} \enspace \left(i=\left\{1,...,N\right\}\right)}$. Additionally, the first invariant of $\underline{\underline{C}}$ is defined as:
\begin{equation}
\label{eq:sm.b1.3}
I_{\mathrm{C}} = \mathrm{tr}\!\left(\underline{\underline{C}}\right) \, ,
\end{equation}
\noindent which represents the sum of the squares of the principal stretches, ${I_{\mathrm{C}} = \sum_{i=1}^{N} \lambda_{i}^{2}}$.  

When modeling a hydrogel as a continuum, the two constituent phases must be incorporated into the kinematics. Because $\underline{u}$ explicitly tracks the deformation of the polymer network, the kinematics of the interstitial water are treated only implicitly. Taking an infinitesimal volume of hydrogel in the current (deformed) configuration $\mathrm{d}V$, we identify the volumes occupied by the polymer and water as $\mathrm{d}V_{\mathrm{p}}$ and $\mathrm{d}V_{\mathrm{w}}$, respectively:
\begin{equation}
\label{eq:sm.b1.4}
\mathrm{d}V = \mathrm{d}V_{\mathrm{p}} + \mathrm{d}V_{\mathrm{w}} \, .
\end{equation}
\noindent Dividing by the corresponding infinitesimal volume of hydrogel in the reference configuration $\mathrm{d}V_{0}$, we obtain:
\begin{equation}
\label{eq:sm.b1.5}
\frac{\mathrm{d}V}{\mathrm{d}V_{0}} = \frac{\mathrm{d}V_{\mathrm{p}}}{\mathrm{d}V_{0}} + \frac{\mathrm{d}V_{\mathrm{w}}}{\mathrm{d}V_{0}} \Rightarrow J = \Phi + \Phi_{\mathrm{w}} \, ,
\end{equation}
\noindent where $\Phi$ and $\Phi_{\mathrm{w}}$ are the nominal (Lagrangian) volume fractions of the polymer and water, respectively. We can equivalently define these magnitudes based on the true (Eulerian) polymer volume fraction ${\phi = \mathrm{d}V_{\mathrm{p}}/\mathrm{d}V}$ as well:
\begin{equation}
\label{eq:sm.b1.6}
\Phi = \phi\, J \quad \mathrm{and} \quad \Phi_{\mathrm{w}} = \left(1-\phi\right) J \, .
\end{equation}

Furthermore, assuming that the hydrogel is intrinsically incompressible, mass conservation dictates that any change in the volume of a Lagrangian element arises solely from changes in water concentration. Therefore, the amount of polymer within the Lagrangian reference volume ${\mathrm{d}V_{0}}$ remains constant, meaning ${\Phi =\phi_{0}}$ regardless of deformation. This leads to a local kinematic constraint linking $\phi$ and $J$: 
\begin{equation}
\label{eq:sm.b1.7}
J = \frac{\phi_{0}}{\phi} \, ,
\end{equation}

\noindent which, in turn, yields the kinematic definition of the Lagrangian volume fractions of polymer $\Phi$ and water $\Phi_{\mathrm{w}}$:
\begin{equation}
\label{eq:sm.b1.8}
\Phi = \phi_{0} \quad \mathrm{and} \quad \Phi_{\mathrm{w}} = J - \phi_{0} \, .
\end{equation}
\subsection{B.2. Thermodynamic constitutive formulation}

In line with the conventional Flory-Rehner modeling paradigm for hydrogels~\cite{Flory1943}, we assume that the total Helmholtz free energy is decomposed additively into an elastic term accounting for polymer network deformation and an (external) mixing term reflecting polymer-water affinity. Mathematically, we write the Lagrangian free energy density $\Psi$ in terms of the elastic ($\Psi_{\mathrm{el}}$) and mixing ($\Psi_{\mathrm{mix}}$) contributions as:
\begin{equation}
\label{eq:sm.b2.1}
\Psi\!\left(\underline{\underline{F}},\, \Phi_{\mathrm{w}}, \, T\right) = \Psi_{\mathrm{el}}(\underline{\underline{F}}) - \Psi_{\mathrm{mix}}\!\left( \Phi_{\mathrm{w}}, \, T \right) \, ,
\end{equation}
\noindent where we assume that $\Psi_{\mathrm{el}}$ depends solely on the deformation gradient $\underline{\underline{F}}$, while $\Psi_{\mathrm{mix}}$ depends on the reference water volume fraction $\Phi_{\mathrm{w}}$ and the absolute temperature $T$. At this stage, the volumetric relation between $\Phi_{\mathrm{w}}$ and $J$ in Eq.~\eqref{eq:sm.b1.8} is not yet considered; hence, $\underline{\underline{F}}$ and $\Phi_{\mathrm{w}}$ remain independent state variables. Alternative formulations~\cite{Hong2008, Bertrand2016, Garyfallogiannis2022} use the Lagrangian molecular concentration of water $C_{\mathrm{w}}$ instead of $\Phi_{\mathrm{w}}$. Yet, assuming a constant molecular volume of water $v_{\mathrm{w}}$ renders these quantities directly proportional:
\begin{equation}
\label{eq:sm.b2.2}
\Phi_{\mathrm{w}} = C_{\mathrm{w}} v_{\mathrm{w}}\,.
\end{equation}

Although $\Phi_{\mathrm{w}}$ (or $C_{\mathrm{w}}$) is a primary thermodynamic state variable for isolated hydrogel systems, it is mechanically inconvenient: environmental conditions typically dictate the chemical potential of water at the hydrogel boundaries rather than its volume fraction (or concentration). Defining $\mu_{\mathrm{w}}$ as the chemical potential of water inside the hydrogel relative to pure water at zero pressure, we can write:
\begin{equation}
\label{eq:sm.b2.3}
\mu_{\mathrm{w}} (\underline{\underline{F}},\, \Phi_{\mathrm{w}},\, T) = \frac{\partial \Psi(\underline{\underline{F}},\, \Phi_{\mathrm{w}},\, T)}{\partial \Phi_{\mathrm{w}}} \, v_{\mathrm{w}} \, .
\end{equation}
Therefore, $\mu_{\mathrm{w}}$ is proportional to the thermodynamic driving force conjugate to $\Phi_{\mathrm{w}}$, which is formally defined as the pore pressure of interstitial water (hereafter, simply pore pressure):
\begin{equation}
\label{eq:sm.b2.4}
p\,(\underline{\underline{F}},\, \Phi_{\mathrm{w}}, \, T) = \frac{\partial \Psi(\underline{\underline{F}},\, \Phi_{\mathrm{w}}, \, T)}{\partial \Phi_{\mathrm{w}}}\, .
\end{equation}

Because environmental boundary conditions on $\mu_{\mathrm{w}}$ directly translate to $p$, and $p$ directly contributes to the continuum stress state, adopting it as the independent thermodynamic variable instead of $\Phi_{\mathrm{w}}$ is highly advantageous. We achieve this change of independent variables via a Legendre transform:
\begin{equation}
\label{eq:sm.b2.5}
\Psi^{*}\!\left(\underline{\underline{F}},\, p,\, T\right) = \inf_{\Phi_{\mathrm{w}}} \left[\Psi\!\left(\underline{\underline{F}},\, \Phi_{\mathrm{w}},\, T\right) - \Phi_{\mathrm{w}} \, p\right] \, ,
\end{equation}
\noindent which yields the Lagrangian grand potential density $\Psi^{*}$ as an explicit function of $(\underline{\underline{F}},\, p,\, T)$. This infimum is reached when the bracketed term is stationary with respect to $\Phi_{\mathrm{w}}$, requiring:
\begin{equation}
\label{eq:sm.b2.6}
\frac{\partial \Psi (\underline{\underline{F}},\, \Phi_{\mathrm{w}},\, T)}{\partial \Phi_{\mathrm{w}}} - p = 0\, .
\end{equation}
This recovers the conjugacy relation from Eq.~\eqref{eq:sm.b2.4} and implicitly defines $\Phi_{\mathrm{w}}$ as a function of $(\underline{\underline{F}}, \, p,\, T)$. Eventually, the grand potential $\Psi^{*}$ takes the general form:
\begin{equation}
\label{eq:sm.b2.7}
\Psi^{*}\!\left(\underline{\underline{F}},\, p,\, T\right) = \Psi\!\left(\underline{\underline{F}},\, \Phi_{\mathrm{w}}(\underline{\underline{F}}, \, p,\, T),\, T\right) - \Phi_{\mathrm{w}}(\underline{\underline{F}}, \, p,\, T) \,\, p \, .
\end{equation}

Thus far, this thermodynamic formulation has not invoked mass conservation. However, recalling the kinematic constraint in Eq.~\eqref{eq:sm.b1.8}, we can explicitly replace $\Phi_{\mathrm{w}}$ with ${J - \phi_{0}}$. Incorporating this functional dependency alongside the energy decomposition in Eq.~\eqref{eq:sm.b2.1}, the grand potential for the mass-conserving hydrogel system can be written as:
\begin{equation}
\label{eq:sm.b2.8}
\Psi^{*}(\underline{\underline{F}},\, p,\, T) = \Psi_{\mathrm{el}}(\underline{\underline{F}}) - \Psi_{\mathrm{mix}}(J,\, T) - \left(J-\phi_{0}\right) \, p \, .
\end{equation}

For the elastic strain energy density~$\Psi_{\mathrm{el}}$, we employ a modified compressible neo-Hookean model~\cite{Garyfallogiannis2022} referenced to the undeformed, fully swollen state: 
\begin{equation}
\label{eq:sm.b2.9}
\Psi_{\mathrm{el}}\!\left(\underline{\underline{F}}\right) = 
\frac{G_{0}}{2} \left[I_{\mathrm{C}} - 3 - 2 \,\, \ln J\right] + \sigma_{0}\!\left(J-1\right) \, ,
\end{equation}
\noindent where $G_{0}$ is the deviatoric modulus in the undeformed configuration. Comparing with the classical Flory-Rehner theory, $G_{0}$ serves as the continuum counterpart of ${N k_{\mathrm{B}} T}$, with $N$ being the chain number density and $k_{\mathrm{B}}$ the Boltzmann constant. Accounting for the difference in reference states —$G_{0}$ is referenced to ${\phi = \phi_{0}}$, while Flory-Rehner elasticity is referenced to $\phi = 1$— we can write that ${G_{0} = N k_{\mathrm{B}} T \phi_{0}^{1/3}}$. The linear volumetric term ${\sigma_{0}\!\left(J-1\right)}$ augments the standard neo-Hookean model by introducing a constant true stress $\sigma_{0}\underline{\underline{I}}$ (see Eq.~\eqref{eq:2} in the main text). This hydrostatic term balances the non-zero mixing pressure at the reference state (${\sigma_{0} = \Pi_\mathrm{mix}(\phi_{0}) \neq 0}$), ensuring a stress-free undeformed configuration (${\underline{\underline{\sigma}}(\underline{\underline{F}} = \underline{\underline{I}}) = \underline{\underline{0}}}$). Because hydrogels are typically highly hydrated at ${p=0}$, their mixing pressure at ${\phi = \phi_{0}}$ is usually low. Consequently, $\sigma_{0}$ is also small; e.g., for PEGDA at ${T = 298.15~\mathrm{K}}$, ${\sigma_{0} = 0.095~\mathrm{MPa}}$.

For the mixing energy density $\Psi_{\mathrm{mix}}$, we adopt a Des Cloizeaux-type power law scaling~\cite{Feng2025}. Recasting its native Eulerian form (in terms of $\phi$) into our Lagrangian formulation via ${\phi = \phi_{0}/J}$ yields:
\begin{equation}
\label{eq:sm.b2.10}
\Psi_{\mathrm{mix}}\!\left(J,\, T\right) = -\frac{A k_{\mathrm{B}}T \phi_{0}^{n}}{v_{w} \left(n-1\right)} \frac{1}{J^{n-1}} \, .
\end{equation}
\noindent where $A$ is a dimensionless energy parameter governing the magnitude of polymer-water affinity, and $n$ is the scaling exponent related to the fractal porous structure of the polymer network ($n>1$). This mixing energy is referenced to the limit of infinite hydration, such that $\Psi_{\mathrm{mix}}(J \rightarrow \infty,\, T<\infty) = 0$.
\subsection{B.3. Mechanical constitutive formulation}

Based on the Lagrangian grand potential~$\Psi^{*}$, the first Piola-Kirchhoff stress tensor $\underline{\underline{P}}$ is defined as the work-conjugate to the deformation gradient~$\underline{\underline{F}}$. Because $\underline{\underline{P}}$ characterizes the purely mechanical response of the system upon deformation at a given pore pressure $p$ and temperature $T$, we write:
\begin{equation}
\label{eq:sm.b3.1}
\underline{\underline{P}}(\underline{\underline{F}};\, p,\, T) = \frac{\partial \Psi^{*}(\underline{\underline{F}};\, p,\, T)}{\partial \underline{\underline{F}}}
\end{equation}

\noindent where the semicolon separates variables (left) from parameters (right). Mechanically, $\underline{\underline{P}}$ represents the nominal stress, i.e. the force transmitted per unit area in the reference configuration. The additive decomposition of $\Psi^{*}$ is preserved in $\underline{\underline{P}}$:
\begin{equation}
\label{eq:sm.b3.2}
\underline{\underline{P}} (\underline{\underline{F}};\, p,\, T) = 
\underline{\underline{P}}_{\mathrm{el}}\!(\underline{\underline{F}}) -
\underline{\underline{P}}_{\mathrm{mix}}\!(\underline{\underline{F}};\, T) -
\underline{\underline{P}}_{\mathrm{p}}\!\!\left(\underline{\underline{F}};\, p\right)\, .
\end{equation}

\noindent where the nominal elastic stress is:
\begin{equation}
\label{eq:sm.b3.3}
\underline{\underline{P}}_{\mathrm{el}}\!(\underline{\underline{F}}) =
G_{0} \! \left(\underline{\underline{F}} - \underline{\underline{F}}^{-\mathrm{T}}\right) + \sigma_{0}\,J \underline{\underline{F}}^{-\mathrm{T}}\, ,
\end{equation}

\noindent the nominal stress due to mixing pressure is:
\begin{equation}
\label{eq:sm.b3.4}
\underline{\underline{P}}_{\mathrm{mix}}\!(\underline{\underline{F}};\, T) = 
\frac{A k_{\mathrm{B}} T}{v_{w}} \frac{\phi_{0}^{n}}{J^{n-1}} \underline{\underline{F}}^{-\mathrm{T}} \, ,
\end{equation}
\noindent and the nominal stress due to pore pressure is:
\begin{equation}
\label{eq:sm.b3.5}
\underline{\underline{P}}_{\mathrm{p}}\!(\underline{\underline{F}};\, p) = 
p \,J \underline{\underline{F}}^{-\mathrm{T}} \, .
\end{equation}

To obtain the true (Cauchy) stress tensor $\underline{\underline{\sigma}}$, we map the reference area elements to the current configuration via Nanson's formula. This yields: 
\begin{equation}
\label{eq:sm.b3.6}
\underline{\underline{\sigma}}(\underline{\underline{F}};\, p,\, T) = \frac{1}{J} \, \underline{\underline{P}}(\underline{\underline{F}};\, p,\, T) \, \underline{\underline{F}}^{\mathrm{T}} \, .
\end{equation}

Because we assume known isothermal conditions throughout this work, we drop the explicit dependence on $T$ in the main text for conciseness. Likewise, we forfeit the semicolon notation. Substituting the nominal stress components in Eqs.~(\mbox{\ref{eq:sm.b3.3}--\ref{eq:sm.b3.5}}) into Eq.\eqref{eq:sm.b3.6} recovers the Cauchy stress definitions provided in the main text (see Eqs.~(\mbox{\ref{eq:1}--\ref{eq:3}})), where $\underline{\underline{B}} = \underline{\underline{F}} \, \underline{\underline{F}}^{\mathrm{T}}$ is the left Cauchy-Green tensor. 
\subsection{B.4. Lagrangian strong form of the quasi-static mechanical problem}
Let us now formally define the quasi-static mechanical problem within this Lagrangian poroelastic framework. Consider a hydrogel body occupying the reference domain $\Omega_{0}$, with an external boundary $\partial\Omega_{0}$ and a unit outward normal $\underline{N}$. The boundary is partitioned into two disjoint subsets: the Dirichlet boundary $\partial_{u}\Omega_{0}$, where displacements are prescribed, and the Neumann boundary $\partial_{\sigma}\Omega_{0}$, where tractions are applied. Neglecting inertial effects and assuming given fields for the body forces $\underline{B}$, pore pressure $p$, and temperature $T$, the strong form of the boundary-value problem governing mechanical equilibrium requires finding the displacement field ${\underline{u}:\Omega_{0} \rightarrow \mathbb{R}^{N}}$ that satisfies:
\begin{equation}
\label{eq:sm.b4.1}
\left\{
\begin{array}{cl}
     \nabla \cdot \underline{\underline{P}}(\underline{\underline{F}};\, p,\, T) + \underline{B} = \underline{0} &  \forall \underline{X} \in \Omega_{0} \\
     \underline{u} = \underline{U} & \forall \underline{X} \in \partial_{u}\Omega_{0} \\
     \underline{\underline{P}} \, \underline{N} = \underline{S} &  \forall \underline{X} \in \partial_{\sigma}\Omega_{0} 
\end{array}
\right.
\end{equation}
\noindent where $\nabla\cdot\underline{\underline{P}}$ is the Lagrangian divergence of $\underline{\underline{P}}$, and $\underline{U}$ and $\underline{S}$ are the prescribed displacement and nominal traction vector fields, respectively. This formulation strictly describes the stationary limit of the poroelastic problem, treating both $p$ and $T$ as known steady-state fields. Solving the fully coupled thermo-hydro-mechanical problem, which requires augmenting mechanical equilibrium with principles of water migration kinetics and thermal diffusion, lies beyond the scope of this work. %
\subsection{B.5. Predicted one-dimensional compressive modulus $M_{\mathrm{1D}}$ under partial dehydration ($\phi \ge \phi_{0}$)}

To provide the predicted one-dimensional compressive modulus $M_{\mathrm{1D}}$ as a function of $\phi$, we replicate the experimental conditions of the gel freezing osmometry test (see Sec.~A.1). Under one-dimensional kinematics governed by the stretch ${\lambda}$, the finite deformation tensor, volume change, and left Cauchy-Green tensor reduce to the scalars ${F = \lambda}$, ${J = \lambda}$, and ${B = \lambda^{2}}$, respectively. The corresponding true elastic stress then simplifies to a scalar:
\begin{equation}
\label{eq:sm.b5.1}
\sigma_{\mathrm{el}} \! \left(\lambda\right) = G_{0}\frac{\lambda^{2} - 1}{\lambda} + \sigma_{0}~ ,
\end{equation}
\noindent and the true mixing pressure becomes:
\begin{equation}
\label{eq:sm.b5.2}
\Pi_{\mathrm{mix}} \! \left(\lambda,\, T\right) = \frac{A k_{\mathrm{B}} T}{v_{\mathrm{w}}} \left(\frac{\phi_{0}}{\lambda}\right)^{n} ~ ,
\end{equation}
leaving the total true stress as:
\begin{equation}
\label{eq:sm.b5.3}
\sigma(\lambda; \, p,\, T) = \sigma_{\mathrm{el}}(\lambda) - \Pi_{\mathrm{mix}} \! \left(\lambda;\, T\right) - p  ~ .
\end{equation}

Under free shrinkage conditions, mechanical equilibrium requires the total true stress to vanish throughout the domain, i.e. ${\sigma = 0}$. This allows us to express the pore pressure $p$ explicitly as a function of $\lambda$ and $T$:
\begin{equation}
\label{eq:sm.b5.4}
p\left(\lambda,\, T\right) = \sigma_{\mathrm{el}} \! \left(\lambda\right) - \Pi_{\mathrm{mix}} \! \left(\lambda,\, T\right) \, .
\end{equation}
\noindent Substituting this pressure into the definition of the one-dimensional compressive modulus (Eq.~\eqref{eq:sm.a1.2}) yields:
\begin{equation}
\label{eq:sm.b5.5}
M_{\mathrm{1D}} \! \left(\lambda,\, T\right) = G_{0} \frac{1 + \lambda^{2}}{\lambda} + 
\frac{A k_{\mathrm{B}} T n}{v_{\mathrm{w}}} \left(\frac{\phi_{0}}{\lambda}\right)^{n} \, .
\end{equation}
\noindent Finally, expressing this in terms of the true polymer volume fraction ${\phi = \phi_{0}/\lambda}$, we obtain:
\begin{equation}
\label{eq:sm.b5.6}
M_{\mathrm{1D}} \! \left(\phi,\, T\right) = G_{0} \! \left( \frac{\phi}{\phi_{0}} + \frac{\phi_{0}}{\phi} \right) + 
\frac{A k_{\mathrm{B}} T \, n}{v_{\mathrm{w}}} \phi^{n} \, .
\end{equation}
\noindent Evaluating this expression with the material properties of PEGDA and the corresponding hydrogel temperature $T$ from the experiments yields the dashed curve reported in Fig.~\ref{fig:1}. 

\subsection{B.6. Predicted apparent shear modulus $G$ under partial dehydration ($\phi \ge \phi_{0}$)}

To provide the predicted apparent shear modulus as a function of dehydration, we bypass the complexity of the micro-indentation experiments (Sec.~A.2) in favor of a simpler, yet physically analogous deformation mode: simple shear under partial dehydration. In this virtual test, we first subject the domain to volumetric shrinkage, defined by:  
\begin{equation}
\label{eq:sm.b6.1}
\underline{\underline{F}}_{\mathrm{vol}} = \lambda_{\mathrm{vol}} \, \underline{\underline{I}} \, ,
\end{equation}
\noindent and then subject it to volume-conserving simple shear. Defining $\underline{e}_{\mathrm{n}}$ as the normal to the shearing plane and $\underline{e}_{\mathrm{sh}}$ as its loading direction (with ${\underline{e}_{\mathrm{n}} \cdot\underline{e}_{\mathrm{sh}} = 0}$), the shearing finite deformation tensor is:
\begin{equation}
\label{eq:sm.b6.2}
\underline{\underline{F}}_{\mathrm{sh}} = \underline{\underline{I}} + \gamma_{\mathrm{sh}} \,\, \underline{e}_{\mathrm{sh}} \! \otimes \underline{e}_{\mathrm{n}} \, .
\end{equation}
\noindent The overall deformation gradient is then obtained via the multiplicative composition ${\underline{\underline{F}}=\underline{\underline{F}}_{\mathrm{vol}} \, \underline{\underline{F}}_{\mathrm{sh}}}$. Treating volumetric shrinkage ${J = \lambda_{\mathrm{vol}}^{3} = \phi_{0} / \phi}$ as the control parameter for dehydration, we apply simple shear as a proxy for localized indentation. Because this isochoric deformation mode is resisted only by the neo-Hookean elastic stress in the strongly coupled formulation, the relevant shear component of the total stress, $\sigma_{\mathrm{sh}}$, is given by: 
\begin{equation}
\label{eq:sm.b6.3}
\sigma_{\mathrm{sh}} = 
\underline{e}_{\mathrm{sh}} \cdot \underline{\underline{\sigma}}_{\mathrm{el}} (\underline{\underline{F}}) \,\, \underline{e}_{\mathrm{n}} = \frac{G_{0}}{\lambda_{\mathrm{vol}}} \, \gamma_{\mathrm{sh}} \, .
\end{equation}

Evaluating the apparent shear modulus $G$ at a given shrinkage $J$ in the small-strain limit ($\gamma_{\mathrm{sh}} \rightarrow 0$) yields:
\begin{equation}
\label{eq:sm.b6.4}
G (J) = \lim_{\gamma_{\mathrm{sh}} \rightarrow 0} \! \left( \frac{\partial \sigma_{\mathrm{sh}}}{\partial \gamma_{\mathrm{sh}}} \right)  = \frac{G_{0}}{J^{1/3}} \, ,
\end{equation}
\noindent Finally, expressing $G$ in terms of the true polymer volume fraction ${\phi = \phi_{0}/J}$, we obtain:
\begin{equation}
\label{eq:sm.b6.5}
G \! \left( \phi \right) = G_{0}\!\left(\frac{\phi}{\phi_{0}}\right)^{\!1/3}  .
\end{equation}
\noindent Evaluating this expression with the material properties of PEGDA yields the dash-dotted curve reported in Fig.~\ref{fig:1}.
\subsection{B.7. Material characterization for PEGDA}

The presented poroelastic model depends on four material properties: the initial polymer volume fraction $\phi_{0}$, the reference deviatoric modulus $G_{0}$, and the mixing parameters $A$ and $n$. For PEGDA, the initial volume fraction is fixed at ${\phi_{0} = 0.2}$. We determine $G_{0}$ by fitting Eq.~\eqref{eq:sm.b6.5} to experimental measurements of ${G\!\left(\phi\right)}$ from the partially dried micro-indentation tests. We employ a weighted non-linear least-squares regression using the \texttt{scipy.optimize.curve\_fit()} function; to prioritize relative accuracy, we set the weights to the experimental $G$ values themselves. Based on this calibration procedure, we obtain a best-fit reference shear modulus ${G_{0} \approx 0.26 \,\, \mathrm{MPa}}$. The comparison between the predicted and experimental values of $G\!\left(\phi\right)$ for PEGDA is shown in Fig.~\ref{fig:1}. 

Using this value for $G_{0}$, we proceed to determine $A$ and $n$ from the gel freezing osmometry data~\cite{Feng2025}. To avoid numerical errors inherent in approximating derivatives from discrete data points, we fit Eq.~\eqref{eq:sm.b5.4} directly to the pressure-stretch ($p$ versus $\lambda$) data rather than to the modulus data reported in Fig.~\ref{fig:1}; we set ${\sigma_0 = 0}$ for simplicity. Following an analogous weighted least-squares procedure to that used for $G_{0}$, we obtain the best-fitting parameters ${A \approx 1.75}$ and ${n \approx 4.67}$. The resulting model predictions for ${p}$ versus $\lambda$ are plotted against the experimental data in Fig.~\ref{fig:sm.b.1}.

\begin{onecolfig}[t]
  \centering
  \includegraphics[width=\linewidth]{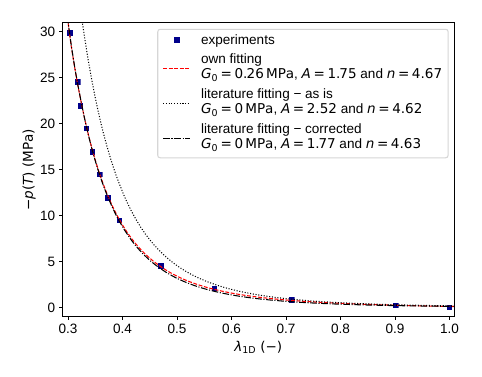}
  \caption{PEGDA model calibration against experimental data and comparison with the literature. The proposed PEGDA model (dashed line) accurately captures the experimental measurements (markers) for pore pressure $p$ versus stretch $\lambda$ data \cite{Feng2025}. A prior model calibration from the same literature source is provided for comparison (dotted line). Notably, the literature model is also plotted using recalculated $A$ values to correct an apparent typographical error in the publication (dashed-dotted line).}
  \label{fig:sm.b.1}
  \vspace{-5mm}
\end{onecolfig}

As a validation step, we compare these calibrated properties against those reported in the relevant literature~\cite{Feng2025}. While minor discrepancies exist, they are readily reconciled by examining the respective fitting strategies. For $G_{0}$, the literature reports a higher value ($G_{0} \approx 0.38\,\,\mathrm{MPa}$) derived exclusively from data at near-full hydration. Restricting our regression to a similarly narrow hydration range (e.g., ${\phi < 1.05\phi_{0}}$) yields virtually the same result. The comparison for $A$ and $n$ is more nuanced. The referenced study neglected the elastic contribution in the gel freezing osmometry test. If we apply the same assumption, namely $G_{0} = 0~\mathrm{MPa}$, our procedure yields ${A \approx 1.77}$ and ${n \approx 4.63}$ compared to the reported ${A \approx 2.52}$ and ${n \approx 4.62}$. While our fitted value for $n$ aligns reasonably well, we note that the literature value for $A$ appears to be a typographical error: it fails to reproduce the reported experimental matching, while our recalculation ${A \approx 1.77}$ does (see Fig.~\ref{fig:sm.b.1}). We therefore conclude that our material characterization procedure is consistent with the relevant literature. Ultimately, the slight variation in the calibrated value for $A$ depending on whether elasticity is considered further underscores the core message of this work: the elastic contribution to volumetric shrinkage in drying hydrogels is negligible.
\subsection{B.8. Uniaxial loading of a 2D representative volume element under controlled pore pressure}

Let us consider a two-dimensional representative volume element within the plane spanning $\underline{e}_{x}$ and $\underline{e}_{y}$. Assuming uniaxial loading along $\underline{e}_{y}$, both $\underline{e}_{x}$ and $\underline{e}_{y}$ act as principal directions, with the corresponding stretches being $\lambda_{x}$ and $\lambda_{y}$. Furthermore, considering plane strain conditions, the out-of-plane deformation in the direction ${\underline{e}_{z} = \underline{e}_{x} \times \underline{e}_{y}}$ is zero; hence, ${\lambda_{z} = 1}$. Consequently, the two-dimensional deformation gradient tensor is defined as:
\begin{equation}
\label{eq:sm.b8.1}
\underline{\underline{F}} = \sum_{i = x, \, y} \!\lambda_{i} \,\, \underline{e}_{i} \otimes \underline{e}_{i} \, ,
\end{equation}
\noindent the corresponding volume change is:
\begin{equation}
\label{eq:sm.b8.2}
J = \lambda_{x}\lambda_{y} \, ,
\end{equation}
\noindent and the left Cauchy-Green tensor is:
\begin{equation}
\label{eq:sm.b8.3}
\underline{\underline{B}} = \sum_{i = x,y} \!\lambda_{i}^{2} \,\, \underline{e}_{i} \otimes \underline{e}_{i} \, .
\end{equation}

Assuming stretch-controlled loading along $\underline{e}_{y}$, stress-free conditions in $\underline{e}_{x}$, and prescribed pore pressure $p$ and temperature $T$, mechanical equilibrium dictates:
\begin{equation}
\label{eq:sm.b8.4}
\left\{ 
\begin{array}{c}
     \underline{e}_{x} \cdot \underline{\underline{\sigma}}(\underline{\underline{F}};\, p,\, T) \, \underline{e}_{x} = 0 \\
     \\
     \underline{e}_{y} \cdot \underline{\underline{\sigma}}(\underline{\underline{F}};\, p, \, T) \, \underline{e}_{y} = \sigma_{yy}
\end{array}
\right. ~.
\end{equation}
\noindent Evaluating the stress along $\underline{e}_{x}$ yields an implicit equation for $\lambda_{x}$ in terms of $\lambda_{y}$, $p$, and $T$:
\begin{equation}
\label{eq:sm.b8.5}
G_{0} \, \frac{\lambda_{x}^{2}-1}{\lambda_{x}\lambda_{y}} -\frac{A \, k_{\mathrm{B}} T}{v_{\mathrm{w}}} \left[\left(\frac{\phi_{0}}{\lambda_{x}\lambda_{y}}\right)^{n} -\phi_{0}^{n}\right] - p = 0 ~.
\end{equation}
\noindent Once $\lambda_{x}$ is determined for the given $\lambda_{y}$, $p$, and $T$, mechanical equilibrium in the loading direction requires that the applied stress $\sigma_{yy}$ satisfies: 
\begin{equation}
\label{eq:sm.b8.6}
\sigma_{yy} = G_{0} \, \frac{\lambda_{y}^{2}-1}{\lambda_{x}\lambda_{y}} -\frac{A \, k_{\mathrm{B}} T}{v_{\mathrm{w}}} \left[\left(\frac{\phi_{0}}{\lambda_{x}\lambda_{y}}\right)^{n} -\phi_{0}^{n}\right] - p ~.
\end{equation}

Evaluating Eqs.~\eqref{eq:sm.b8.5} and \eqref{eq:sm.b8.6} using the calibrated properties of PEGDA yields the plots presented in Fig.~\ref{fig:3}, where we assumed that the hydrogel is at standard room temperature, i.e. $T = 298.15~\mathrm{K}$. 
\section{C. Numerical aspects of the strongly coupled steady-state formulation}

We implement the strongly coupled poroelastic formulation using the open-source finite element (FE) library FEniCSx (\texttt{v0.8.0}). Our workflow begins by generating the geometrically discretized domains via the Python API of Gmsh. We then define the numerical FE problem by discretizing the variational formulation (derived in Sec.~C.1) using second-order continuous Galerkin elements (\texttt{CG2}). Besides improving accuracy compared to first-order elements, this choice maintains coherence in the kinematic discretization with the Taylor-Hood elements used for the weakly coupled formulation (detailed in Sec.~D~and~E), thus ensuring an equitable comparison. As in conventional FE implementations of solid mechanics problems, we take the nodal displacement field as the primary unknown. 

We mitigate convergence issues arising from the highly nonlinear nature of the problem by adopting an incremental loading strategy with zeroth-order continuation. At each load step, we use the \texttt{newtonls} solver from the SNES interface of PETSc, configured with a basic line search strategy. We explicitly set a relative tolerance of $10^{-10}$ on the residual norm for convergence; the other tolerances remain at the PETSc default. At each Newton iteration, we solve the corresponding linearized system via direct LU factorization with the \texttt{MUMPS} backend. Despite its higher computational cost, this choice generally improves robustness compared to iterative Krylov solvers, which may suffer from slow convergence and stagnation under severe ill-conditioning arising from multiscale stiffening. Upon convergence at each load step, we compute the secondary fields derived from displacement gradients, namely volumetric shrinkage and stresses, and project them onto the corresponding FE function spaces. Finally, we export the fields of interest to ParaView for visualization. The following subsections detail the derivation of the (continuum) Lagrangian variational formulation and provide a detailed description of the numerical models used for the case studies. We note that since the FE discretization is handled internally by \mbox{FEniCSx}, we do not distinguish between continuum and discretized fields in the subsequent derivations. 

\subsection{C.1. Lagrangian variational formulation of the quasi-static mechanical problem}

For a displacement field $\underline{u}$ and prescribed pore pressure $p$ and temperature $T$ fields, the total grand potential energy functional $\mathcal{E}^{*}$ is defined as:
\begin{equation}
\label{eq:sm.c1.1}
\mathcal{E}^{*}\!\left(\underline{u}; \, p,\, T\right) =  \int_{\Omega_{0}} \Psi^{*}\!\left(\underline{\underline{F}}; \, p,\, T\right) \, \mathrm{d}\underline{X} \, .
\end{equation}
\noindent The mechanical work done by external body forces $\underline{B}$ and surface tractions $\underline{S}$ is:
\begin{equation}
\label{eq:sm.c1.2}
\mathcal{W}\!\left(\underline{u}\right) =  
\int_{\Omega_{0}} \! \underline{B} \cdot \underline{u} \, \mathrm{d}\underline{X} +
\int_{\partial_{\sigma}\Omega_{0}} \! \underline{S} \cdot \underline{u} \, \mathrm{d}\underline{X} \, ,
\end{equation}

\noindent Therefore, the total potential energy functional of the system, $\mathcal{P}$, is:
\begin{equation}
\label{eq:sm.c1.3}
\mathcal{P}\left(\underline{u}; \, p,\, T\right) = \mathcal{E}^{*}\!\left(\underline{u} ; \, p,\, T\right) - \mathcal{W}\!\left(\underline{u}\right)\, .
\end{equation}

Mechanical equilibrium is reached at the admissible displacement $\underline{u}$ that locally minimizes the potential energy $\mathcal{P}$ for the given fields $p$ and $T$. We define the set of kinematically admissible displacements $\mathrm{V}$ as an affine subspace of the Sobolev space $[H^{1}\!\left(\Omega_{0}\right)]^{N}$ induced by the prescribed Dirichlet boundary conditions. This yields:
\begin{equation}
\label{eq:sm.c1.4}
\mathrm{V} = 
\left\{\underline{u} \in \left[H^{1}\!\left(\Omega_{0}\right)\right]^{N} \, 
\left\vert \,   
     \underline{u} = \underline{U} \quad \forall \underline{X}\in \partial_{u} \Omega_{0} 
 \right.
\right\} \, .
\end{equation}
\noindent Physical admissibility also requires that local orientation is preserved, i.e., $J > 0$; however, we do not enforce this condition upon implementation. Instead, we algorithmically prevent material collapse or inversion (i.e., ${J\le0}$) via the incremental loading strategy. With sufficiently small load steps starting from the reference configuration (${J = 1}$), the singular volumetric terms at ${J=0}$, specifically $\ln(J)$ and $1/J^{n-1}$, act as an energy barrier as $J\rightarrow0^{+}$. This strongly penalizes the loss of local orientation and effectively ensures that the energy density $\Psi^{*}$ remains finite in practice. Given an admissible displacement ${\underline{u} \in \mathrm{V}}$, kinematically admissible perturbations are of the form ${\underline{u}+\delta\underline{u}}$, where ${\delta\underline{u}}$ belongs to $\mathrm{V}_{0}$, the homogeneous counterpart of $\mathrm{V}$. 

In terms of first-order optimality, the variational formulation of the mechanical equilibrium problem (defined in strong form in Sec.~B.4) consists of finding the displacement field $\underline{u}\in\mathrm{V}$ such that:
\begin{equation}
\label{eq:sm.c1.5}
\mathrm{D}_{u} \! \left[ \mathcal{P}\left(\underline{u} ; \, p,\, T\right)\right]\left(\delta\underline{u}\right) = 0 \quad \forall \delta\underline{u} \in \mathrm{V}_{0}
\end{equation}
\noindent where $\mathrm{D}_{u} \left[ \mathcal{P}\left(\underline{u};p,\, T\right)\right]\left(\delta\underline{u}\right)$ denotes the G\^{a}teaux derivative of the functional $\mathcal{P}\left(\underline{u};p,\, T\right)$ with respect to $\underline{u}$ and in the direction of $\delta\underline{u}$. This condition enforces only first-order stationarity of $\mathcal{P}$, which is generally non-convex in $\underline{u}$. Evaluating the second-order optimality conditions is beyond the scope of this work. This aligns with conventional practices in nonlinear hyperelasticity problems, which are formally analogous to our formulation, since both the pore pressure $p$ and temperature $T$ are treated as prescribed fields. Such parallelism allows us to leverage standard computational techniques from hyperelasticity to solve the strongly coupled formulation. 
\subsection{C.2. Details of the numerical implementation}

Our numerical implementation of the strongly coupled formulation comprises two distinct case studies: the annular disk (Fig.~\ref{fig:2} and Fig.~\ref{fig:sm.c.1}a) and the chiral 7-arm spiral (Fig.~\ref{fig:4} and Fig.~\ref{fig:sm.c.1}b). For both case studies, the geometries are two-dimensional and assumed to deform under plane strain conditions. Furthermore, we consider a uniform standard room temperature (${T = 298.15~\mathrm{K}}$), noting that the constitutive temperature dependence is very weak within this practical range. For postprocessing simplicity, nodal results in the second-order Galerkin space ($\texttt{CG2}$) are interpolated into the corresponding first-order counterpart ($\texttt{CG1}$) for both Figs.~\ref{fig:2} and \ref{fig:4}.

For the annular disk, the internal and external radii are set to $2$ and $5~\mathrm{mm}$, respectively. Exploiting the symmetry of both geometry and loading, we model only one quarter of the disk. We mesh this domain using a grid of structured quadrilateral elements, comprising 100 elements in the radial direction and 50 in the hoop direction. The inner boundary is fully constrained in displacement ($\underline{u} = \underline{0}$), while the outer radius is a stress-free boundary. Standard symmetry boundary conditions are applied to the facets intersecting the $X$ and $Y$ axes. Dehydration-driven shrinkage is governed by a homogeneous pore pressure prescribed in the bulk, which is monotonically decreased from ${p = 0~\mathrm{MPa}}$ to ${p = -20~\mathrm{MPa}}$ in increments of $\Delta p = -0.2~\mathrm{MPa}$. If no convergence is reached after a certain time increment with respect to the last converged result, we halve the step size and reattempt solving until successful. 

\begin{onecolfig}[t]
  \centering
  \includegraphics[width=\linewidth]{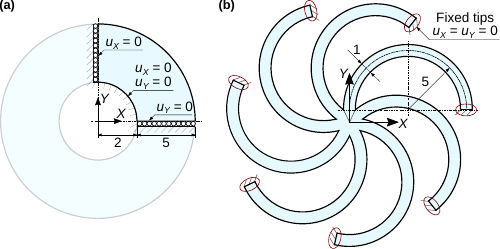}
  \caption{Geometrical description of the reference configuration and applied boundary conditions for (a) the annular, and (b) the spiral numerical case studies. All dimensions are expressed in millimeters. For the spiral, the distal tips are circled to denote that they are all fixed in space (${u_{X} = u_{Y} = 0}$).}
  \label{fig:sm.c.1}
\end{onecolfig}

For the spiral, the individual arms have a midline radius of $5~\mathrm{mm}$ and a uniform thickness of $1~\mathrm{mm}$. The geometry of all the arms is identical, and they are angularly arranged in even intervals of $2\pi/7~\mathrm{rad}$, forming a central hub with a regular heptagonal shape. The domain is meshed using triangular elements of characteristic size $0.05~\mathrm{mm}$, with the discretization strategy varying between regions. While a structured grid is defined for the arms, the central hub is discretized with an unstructured mesh to mitigate severe element distortions arising from radial convergence. The distal flat tips of the arms are all constrained in displacement (${\underline{u} = \underline{0}}$), while the rest of the boundary is stress-free. To induce heterogeneous shrinkage, we prescribe a spatially varying pore pressure field in the bulk. In particular, $p$ is defined by ${p(X) = -(X+10)~\mathrm{MPa}}$, where $X$ is the horizontal material coordinate expressed in $\mathrm{mm}$. Because $p$ is defined in material coordinates, it advects with the deformation of the domain. As in the annular case study, the pore pressure profile is monotonically scaled by a pseudo-time parameter $\tau\in[0,\,1]$, which increments by ${\Delta \tau = 0.01}$ at each load step. Results in the main text correspond only to the final load step, i.e., ${\tau = 1}$. We maintain the semi-adaptive halving strategy for the time-stepping described in the above paragraph.
\section{D. Weakly coupled poroelastic formulation for hydrogel mechanics}

The Lagrangian kinematic framework introduced in Sec.~B.1 remains applicable in the weakly coupled formulation. To maintain a clear distinction between the respective solutions, we now adopt a superimposed hat notation (e.g., $\underline{\hat{u}}$ denotes the weakly coupled solution for the displacement). In contrast to the strongly coupled formulation, where volume and shape are resolved simultaneously, the weakly coupled formulation follows a sequential solution strategy: the local volume change is first determined by the equilibrium between mixing and pore pressures and is then enforced as a volumetric constraint for the elastic shape deformation. Because the weakly coupled formulation only holds for drying hydrogels, we restrict our analysis to negative pore pressure conditions, i.e., ${p\le0}$. Below, we provide the corresponding kinematics, constitutive equations, governing principles, and analytical particularizations. To maintain mathematical rigor, we ensure the exact fulfillment of the volumetric constraint via Lagrange multipliers.

\subsection{D.1. Energy-based constitutive formulation}

Equating the mixing pressure $\Pi_{\mathrm{mix}}$ from Eq.~\eqref{eq:3} to the pore pressure $p$ and inverting the relation yields:
\begin{equation}
\label{eq:sm.d1.1}
\Pi_{\mathrm{mix}}^{-1}\!\left(-p,\, T\right) = \left(-\frac{v_{w}}{A k_{\mathrm{B}} T} \, p\right)^{1/n}\, ,
\end{equation}
\noindent which defines the polymer volume fraction in the zero-elasticity solution (${\phi = \Pi_{\mathrm{mix}}^{-1}(-p)}$) that dominates the volumetric response of drying hydrogels (see Fig.~\ref{fig:2}). However, Eq.~\eqref{eq:sm.d1.1} fails to recover the initial state since ${\lim_{p\rightarrow 0}\Pi_{\mathrm{mix}}^{-1} = 0 \neq \phi_{0}}$. To ensure that the weakly coupled formulation remains physically self-consistent as ${p \rightarrow 0}$, we regularize $\hat{\phi}$ by imposing a threshold at its initial value:
\begin{equation}
\label{eq:sm.d1.2}
\hat{\phi} (p, \,T) = \max\left[\phi_{0},\, \Pi_{\mathrm{mix}}^{-1}\!\left(-p,\, T\right)\right] \, .
\end{equation}
\noindent This correction is only active in the immediate vicinity of $p=0$ (e.g., ${p > -0.13~\mathrm{MPa}}$ for PEGDA at ${T = 298.15~\mathrm{K}}$), where the corresponding near-full hydration already diminishes the validity of the scale separation required for the weakly coupled approximation (see Fig.~\ref{fig:1}). Consequently, the $\mathrm{max}$ operator serves as a pragmatic correction with negligible impact on the intended range of application for the formulation (i.e., ${p<0}$ and ${\phi>\phi_{0}}$). For cases where it is required that $\hat{\phi}(p,\,T)$ has $C^{1}$ continuity or that it is invertible, a mathematically rigorous course of action consists of regularizing the $\mathrm{max}$ operator into the $\mathrm{softmax}$, defined as:
\begin{equation}
\label{eq:sm.d1.3}
\mathrm{softmax}_{\epsilon} \! \left(a, b\right) = \epsilon \, \log\!\left[\mathrm{exp}\!\left({a/\epsilon}\right) + \mathrm{exp}\!\left({b/\epsilon}\right) \right] ~,
\end{equation}
\noindent where $\varepsilon$ is a regularization parameter that governs the abruptness of the transition. In the limit ${\epsilon \rightarrow 0}$, both operators converge. Because the thermodynamics of mixing and ambient drying are now encapsulated in an algebraic constraint (Eqs.~(\mbox{\ref{eq:sm.d1.1}, \ref{eq:sm.d1.2}})), the three-term energetic competition in Sec.~B.2 is effectively bypassed. The weakly coupled formulation can instead be framed using an effective pseudo-energy corresponding to a volume-constrained elastic problem.

For coherence with the strongly coupled formulation, we use a neo-Hookean-type constitutive law for elastic deformation. The thresholding operator in Eq.~\eqref{eq:sm.d1.2} cancels the effect of the non-zero $\Pi_{\mathrm{mix}}$ at ${\hat{J} = 1}$, precluding the need for $\sigma_{0}$ in the elastic energy term. With ${\sigma_{0} = 0}$, the modified neo-Hookean elastic energy density employed in the strongly coupled formulation (Sec.~B.2) reduces to its standard form:
\begin{equation}
\label{eq:sm.d1.4}
\hat{\Psi}_{\mathrm{el}}(\underline{\underline{\hat{F}}}) = 
\frac{G_{0}}{2} \left[\hat{I}_{\mathrm{C}} - 3 - 2 \, \ln\!\hat{J}\right] \, ,
\end{equation}
\noindent which governs the volume-constrained shape deformation. To locally enforce the volume change in an exact manner, we adapt conventional techniques from incompressible solid mechanics and introduce a Lagrange multiplier field $\hat{\pi}$. Physically, $\hat{\pi}$ acts as an internal reaction pressure that locally forces the system to follow the prescribed volume change, such that ${\hat{J} = \phi_{0}/\hat{\phi} (p, \,T)}$. The associated constraint energy density in the reference configuration is:  
\begin{equation}
\label{eq:sm.d1.5}
\hat{\Psi}_{\pi}(\underline{\underline{\hat{F}}},\, \hat{\pi};\, p,\, T) = 
\hat{\pi} \left[\hat{J} - \frac{\phi_{0}}{\hat{\phi} (p, \,T)}\right] \, .
\end{equation}

The sum of elastic and constraint energy contributions yields the augmented energy density for the weakly coupled formulation:
\begin{equation}
\label{eq:sm.d1.6}
\hat{\Psi}(\underline{\underline{\hat{F}}},\, \hat{\pi};\, p,\, T) = 
\hat{\Psi}_{\mathrm{el}}(\underline{\underline{\hat{F}}}) + \hat{\Psi}_{\pi}(\underline{\underline{\hat{F}}},\, \hat{\pi};\, p,\, T)\, .
\end{equation}
\noindent While not a thermodynamic free energy in the strict sense, this constitutes a saddle-point variational functional whose stationarity with respect to $\underline{\hat{u}}$ and $\hat{\pi}$ defines the mechanical equilibrium of the volume-constrained system. This formulation parallels the energetic structure of exactly incompressible hyperelasticity, with the distinction that the local volumetric change is not fixed at unity but is instead a prescribed function of pore pressure and temperature fields.
\subsection{D.2. Mechanical constitutive formulation}

Given that $\hat{\Psi}$ defines the energetic landscape of the weakly coupled formulation, it serves as the counterpart to $\Psi^{*}$ from the strongly coupled formulation. The corresponding first Piola-Kirchhoff stress tensor $\underline{\underline{\hat{P}}}$ is obtained via kinematic differentiation:
\begin{equation}
\label{eq:sm.d2.1}
\underline{\underline{\hat{P}}}(\underline{\underline{\hat{F}}}, \, \hat{\pi}) =
\frac{\partial \hat{\Psi}(\underline{\underline{\hat{F}}}, \, \hat{\pi}; \, p,\, T)}{\partial \underline{\underline{\hat{F}}}}
\, ,
\end{equation}
\noindent where we have already considered that $\underline{\underline{\hat{P}}}$ is inherently independent of $p$ and $T$: these two variables only determine the target volume change ${\phi_{0}/\hat{\phi}(p,\, T)}$. As in the strongly coupled formulation, the total nominal stress $\underline{\underline{\hat{P}}}$ maintains the additive decomposition: 
\begin{equation}
\label{eq:sm.d2.2}
\underline{\underline{\hat{P}}}(\underline{\underline{\hat{F}}}, \, \hat{\pi}) =
\underline{\underline{\hat{P}}}_{\mathrm{el}}\!(\underline{\underline{\hat{F}}}) + \underline{\underline{\hat{P}}}_{\pi}\!(\underline{\underline{\hat{F}}}, \, \hat{\pi})
\, ,
\end{equation}
\noindent with the nominal elastic stress component being:
\begin{equation}
\label{eq:sm.d2.3}
\underline{\underline{\hat{P}}}_{\mathrm{el}}\!(\underline{\underline{\hat{F}}}) =
G_{0} \! \left(\underline{\underline{\hat{F}}} - \underline{\underline{\hat{F}}}^{-\mathrm{T}}\right) \, ,
\end{equation}
\noindent and the nominal volumetric constraint stress component being:
\begin{equation}
\label{eq:sm.d2.4}
\underline{\underline{\hat{P}}}_{\pi}\!(\underline{\underline{\hat{F}}}, \, \hat{\pi}) = \hat{\pi} \hat{J} \, \underline{\underline{\hat{F}}}^{-\mathrm{T}}\, .
\end{equation}

Replicating the push-forward operation from Eq.~\eqref{eq:sm.b3.6}, we recover the total true stress tensor $\underline{\underline{\hat{\sigma}}}$ in Eq.~\eqref{eq:5}.
\subsection{D.3. Lagrangian strong form of the quasi-static mechanical problem}

Adopting the domain and boundary conditions already established in Sec.~B.4, the strong form of the quasi-static mechanical problem for the weakly coupled formulation is defined as a constrained boundary value problem. In particular, it requires finding the fields for the displacement $\underline{\hat{u}}$ and the Lagrange multiplier $\hat{\pi}$ that satisfy:
\begin{equation}
\label{eq:sm.d3.1}
\left\{
\begin{array}{cl}
     \nabla \cdot \underline{\underline{\hat{P}}}(\underline{\underline{\hat{F}}}, \, \hat{\pi}) + \underline{B} = \underline{0} & 
     \forall \underline{X} \in \Omega_{0} \\
     \displaystyle \hat{J} = \frac{\phi_{0}}{\hat{\phi}(p,\, T)} &  \forall \underline{X} \in \Omega_{0} \\
     \underline{\hat{u}} = \underline{U} & \forall \underline{X} \in \partial_{u}\Omega_{0} \\
     \underline{\underline{\hat{P}}} \, \underline{N} = \underline{S} &  \forall \underline{X} \in \partial_{\sigma}\Omega_{0} 
\end{array}
\right.
\end{equation}
\noindent The standard momentum balance and boundary conditions are therefore augmented by the local kinematic constraint acting throughout the domain. Consequently, the paradigm shift in how the problem depends on $p$ and $T$ becomes evident: rather than acting as internal driving forces in the balance of momentum, they now serve as external fields that strictly prescribe the local volumetric change. 
\subsection{D.4. Uniaxial loading of a 2D representative volume element under controlled pore pressure}

Reproducing the 2D plane strain kinematics described in Sec.~B.8, we assume stretch-controlled loading along $\underline{e}_{y}$, free boundaries along $\underline{e}_{x}$, no deformation along $\underline{e}_{z}$, parameterized pore pressure $p$, and prescribed temperature $T = 298.15~\mathrm{K}$. Under these conditions, the mechanical equilibrium for the weakly coupled formulation reduces to the following algebraic system of equations:
\begin{equation}
\label{eq:sm.d4.1}
\left\{ 
\begin{array}{c}
     \underline{e}_{x} \cdot \underline{\underline{\hat{\sigma}}}(\underline{\underline{\hat{F}}},\, \hat{\pi}) \, \underline{e}_{x} = 0 \\
     \\
     \underline{e}_{y} \cdot \underline{\underline{\hat{\sigma}}}(\underline{\underline{\hat{F}}},\, \hat{\pi}) \, \underline{e}_{y} = \hat{\sigma}_{yy} \\
     \\
     \displaystyle\hat{J} = \frac{\phi_{0}}{\hat{\phi}(p,\, T)}
\end{array}
\right. ~ .
\end{equation}

Evaluating the mechanical equilibrium in the $\underline{e}_{x}$ direction while enforcing the volumetric constraint yields a system of two equations for $\hat{\lambda}_{x}$ and $\hat{\pi}$ in terms of $\hat{\lambda}_{y}$, $p$, and $T$:
\begin{equation}
\label{eq:sm.d4.2}
\left\{ 
\begin{array}{c}
     \displaystyle G_{0} \, \frac{\hat{\lambda}_{x}^{2}-1}{\hat{\lambda}_{x}\hat{\lambda}_{y}} + \hat{\pi} = 0 \\
     \\
     \displaystyle \hat{\lambda}_{x} \hat{\lambda}_{y} = \frac{\phi_{0}}{\hat{\phi}(p,\, T)}
\end{array}
\right. ~ ,
\end{equation}
\noindent which admits an explicit analytical solution for $\hat{\lambda}_{x}$ and $\hat{\pi}$:
\begin{equation}
\label{eq:sm.d4.3}
\left\{ 
\begin{array}{c}
     \displaystyle \hat{\lambda}_{x} = \frac{\phi_{0}}{\hat{\lambda}_{y} \, \hat{\phi}(p,\,T)} \\
     \\
     \displaystyle \hat{\pi} = -G_{0} \frac{ \phi_{0}^{2} - \left[\hat{\lambda}_{y}\, \hat{\phi}(p,\, T) \right]^{2}}{\hat{\lambda}_{y} \, \phi_{0} \, \hat{\phi}(p,\,T)}
\end{array}
\right. ~ .
\end{equation}
\noindent The mathematical simplicity of the weakly coupled formulation contrasts with the iterative root finder required for the strongly coupled formulation (Sec.~B.8). Once $\hat{\lambda}_{x}$ and $\hat{\pi}$ are known, mechanical equilibrium along $\underline{e}_{y}$ yields:
\begin{equation}
\label{eq:sm.d4.4}
\hat{\sigma}_{yy} = \displaystyle G_{0} \, \frac{\hat{\lambda}_{y}^{2}-1}{\hat{\lambda}_{x}\hat{\lambda}_{y}} + \hat{\pi}
\end{equation}
\noindent which, together with Eq.~\eqref{eq:sm.d4.3}, yields the results in Fig.~\ref{fig:3} of the main text after considering the properties of PEGDA and the same applied stretch, i.e., ${\lambda_{y} = \hat{\lambda}_{y}}$. 

These analytical solutions directly reveal why the weakly coupled formulation ceases to be accurate under combined severe uniaxial compression (${\hat{\lambda}_{y} \rightarrow 0}$) and vanishing pore pressure (${p \rightarrow 0}$). The isolated polymer network acts as a highly compliant, sponge-like matrix with an effectively zero Poisson's ratio: longitudinal loading yields no transverse deformation when left laterally unconstrained. However, the weakly coupled formulation links $\hat{\lambda}_{x}$ and $\hat{\lambda}_{y}$ through the volumetric constraint, enforcing ${\hat{\lambda}_{x} = \phi_{0}/(\hat{\lambda}_{y} \hat{\phi}(p,\,T))}$ via $\hat{\pi}$. Consequently, as ${p \rightarrow 0}$, the domain must transversely stretch to ${\hat{\lambda}_{x} = 1/\hat{\lambda}_{y}}$, which diverges in the ${\hat{\lambda}_{y} \rightarrow 0}$ limit. Because elastic stresses scale quadratically with stretch, complying with the volumetric constraint generates massive elastic resistance despite the network's high compliance. By transverse mechanical equilibrium (see Eq.~\eqref{eq:sm.d4.2}), this large elastic stress must be countered by an equally large $\hat{\pi}$. This violates the separation of scales required for the weakly coupled formulation to be accurate. An analogous divergence occurs with large tensile elongations (${\hat{\lambda}_{y} \rightarrow \infty}$) and vanishing pore pressure. However, and crucially, because ${\Pi_{\mathrm{mix}}}$ rapidly outgrows $\hat{\pi}$ when ${p<0}$ induces shrinkage, the extent of this inaccurate region is limited. 

\subsection{D.5. Self-contained accuracy metric}

\begin{onecolfig}[t]
  \centering
  \includegraphics[width=\linewidth]{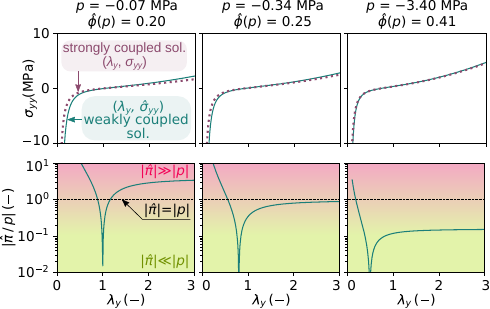}
  \caption{The accuracy of the weakly coupled formulation for a drying hydrogel can be inferred from the ${\vert \hat{\pi} / p \vert \ll 1}$ condition. Upper row of subplots reproduces the stress--stretch curves at different degrees of dehydration from Fig.~\ref{fig:3}; lower row reports the corresponding values of ${\vert \hat{\pi} / p \vert}$.}
  \label{fig:app.1}
  \vspace{-5mm}
\end{onecolfig}

We postulate that ${\vert \hat{\pi}/p \vert}$ is a self-contained metric for assessing the accuracy of the weakly coupled formulation. In the strongly coupled formulation, applied mechanical loads are balanced by a combination of elastic stress and mixing and pore pressures (see Eq.~\eqref{eq:1}). However, the separation of scales in stiffness often leads to the latter two terms effectively canceling each other out in this balance (${\Pi_{\mathrm{mix}} \approx -p}$); thus, the elastic stress is largely left to counter the mechanical loading alone. The weakly coupled formulation makes this cancellation exact (${\Pi_{\mathrm{mix}} = -p}$), enforcing the resulting volumetric constraint on the elastic problem through the Lagrange multiplier $\hat{\pi}$ (see Eq.~\eqref{eq:5}). Consequently, $\hat{\pi}$ acts as a mechanical corrector: it quantifies the extent to which the elastic deformation must be adjusted to satisfy ${\Pi_{\mathrm{mix}} = -p}$. If the ratio ${\vert \hat{\pi}/p \vert}$ is small, the purely elastic solution naturally tends to recover the volumetric constraint associated with ${\Pi_{\mathrm{mix}} = -p}$. In the ${\vert \hat{\pi}/p \vert \rightarrow 0}$ limit, the weakly coupled solution tends towards the strongly coupled solution. Crucially, both $\hat{\pi}$ and $p$ are inherent quantities of the weakly coupled formulation. Therefore, we can evaluate its accuracy without solving the strongly coupled problem for comparison. Reproducing \mbox{Fig.~\ref{fig:3}b.1} and comparing these plots against ${\vert \hat{\pi}/p \vert}$ confirms our premise: the discrepancy between the strongly and weakly coupled formulations is only relevant when the ratio ${\vert \hat{\pi}/p \vert}$ ceases to be small (see Fig.~\ref{fig:app.1}).
\section{E. Numerical aspects of the weakly coupled steady-state poroelastic model}

We implement the weakly coupled formulation using the same computational framework described in Sec.~C. Since the volumetric subproblem reduces to a local algebraic relation (see Sec.~D), only the shape subproblem requires numerical resolution. We define the corresponding finite element problem by discretizing the mixed-space variational formulation (see Sec.~E.1). Because this problem exhibits a saddle-point-like structure, we choose Taylor--Hood elements to ensure inf-sup stability. Thus, we use second-order (\texttt{CG2}) and first-order continuous Galerkin (\texttt{CG1}) discretizations for the displacement $\underline{\hat{u}}$ and Lagrange multiplier $\hat{\pi}$ fields, respectively. To enable a fair comparison of computational performance with the strongly coupled formulation, we retain the same solver configuration as in Sec.~C, namely a direct \texttt{LU} factorization using the \texttt{MUMPS} backend. We note that this choice is suboptimal for mixed formulations, as it does not exploit their inherent block structure. More efficient alternatives exist, such as Schur complement-based block preconditioners and iterative solvers; however, these advanced computational considerations are beyond the scope of this work. We numerically implement the weakly coupled formulation only for the spiral case, using the same setup as in Sec.~C.2 for consistency. The following subsections detail the derivation of the (continuum) Lagrangian variational formulation and discuss the computational performance of the strongly and weakly coupled formulations. 

\subsection{E.1. Lagrangian variational formulation of the shape subproblem}

Following the energy-based constitutive framework detailed in Sec.~D.1, we define the effective augmented energy functional $\hat{\mathcal{E}}$ as:
\begin{equation}
\label{eq:sm.e1.1}
\hat{\mathcal{E}}\!\left(\underline{\hat{u}},\, \hat{\pi};\, p,\, T\right) = \int_{\Omega_{0}} \!\hat{\Psi} (\underline{\underline{\hat{F}}},\, \hat{\pi};\, p,\, T) \, \mathrm{d}\underline{X} ~ ,
\end{equation}
\noindent which acts as an internal pseudo-energy for the hydrogel that deforms according to the prescribed volumetric constraint. The mechanical work done by the external body and surface forces $\mathcal{W}_{\mathrm{ext}}$ remains analogous to that in the strongly coupled formulation (see Sec. B.2). Therefore, we define the effective potential energy $\hat{\mathcal{P}}$ governing shape deformation according to the weakly coupled formulation as:
\begin{equation}
\label{eq:sm.e1.2}
\hat{\mathcal{P}}\left(\underline{\hat{u}}, \, \hat{\pi} ; \, p, \, T\right) = \hat{\mathcal{E}}\!\left(\underline{\hat{u}}, \, \hat{\pi}  ; \, p,\, T\right) - \mathcal{W}_{\mathrm{ext}}\!\left(\underline{\hat{u}}\right) ~ .
\end{equation}

Unlike the strongly coupled formulation, which has displacement as its sole primary variable, this formulation introduces the Lagrange multiplier as an additional independent field. The resulting volume-constraint problem endows the mechanical equilibrium principle with a saddle-point structure, where admissible states ${(\underline{\hat{u}}\in \hat{\mathrm{V}},\, \hat{\pi} \in \hat{\mathrm{Q}})}$ correspond to stationary points of the effective potential energy $\hat{\mathcal{P}}$. Stationarity with respect to $\underline{\hat{u}}$ and $\hat{\pi}$ yields the balance of linear momentum and enforces the volumetric constraint, respectively. The displacement admissibility space $\hat{\mathrm{V}}$ coincides with $\mathrm{V}$ described in Sec.~C.1. Conversely, because $\hat{\Psi}$ depends only on the field $\hat{\pi}$ and not on its spatial derivatives, no gradient regularity is required for its admissibility. The associated admissibility space $\hat{\mathrm{Q}}$ is therefore defined as:  
\begin{equation}
\label{eq:sm.e1.3}
\hat{\mathrm{Q}} = \left\{\hat{\pi} \in L^{2}(\Omega_{0}) \right\} ~,
\end{equation}
\noindent meaning that $\hat{\pi}$ is a scalar field square-integrable on $\Omega_{0}$. In terms of first-order optimality, saddle points require the energy functional $\hat{\mathcal{P}}$ to be stationary with respect to both $\underline{\hat{u}}$ and $\hat{\pi}$; hence:
\begin{equation}
\label{eq:sm.e1.4}
\begin{array}{cc}
\mathrm{D}_{\hat{u}}\!\left[\hat{\mathcal{P}}\left(\underline{\hat{u}}, \, \hat{\pi} ; \, p,\, T\right)\right]\left(\delta\underline{\hat{u}}\right) + 
\mathrm{D}_{\hat{\pi}}\!\left[\hat{\mathcal{P}}\left(\underline{\hat{u}}, \, \hat{\pi} ; \, p,\, T\right)\right]\left(\delta\hat{\pi}\right) = 0 \\
\forall \left(\delta\underline{\hat{u}},\, \delta\hat{\pi}\right) \in \hat{\mathrm{V}}_{0} \times \hat{\mathrm{Q}} ~ ,
\end{array}
\end{equation}
\noindent which represents the variational formulation implemented in our finite element framework. Because ${\delta\underline{\hat{u}}}$ and ${\delta\hat{\pi}}$ are independent in a variational sense, this stationarity condition decomposes into two separate Euler--Lagrange equations. As stated in Sec.~C.1, considering second-order optimality conditions falls outside the scope of this work. Likewise, this formulation is mathematically analogous to that of exactly incompressible elasticity~\cite{Holzapfel2000}, once again highlighting the direct applicability of well-established knowledge from incompressible solids.
%

%
%
%

%

%

%

%
%
%
%
%
%

%

%

%
%
%
%
%
%

%
%
%

%
%
%
%
%
%
%
%
%
%
%
%
%
%
%
%
%
%
%
%
%
%
%
%
%
%
%
%
%
%
%
%

\end{document}